\documentclass{iopjournal}
\usepackage{graphicx}
\usepackage{amsmath}
\usepackage[numbers,sort&compress]{natbib}
\usepackage[version=4]{mhchem}
\usepackage{multirow}
\usepackage{subcaption}

\newcommand{\code}[1]{{\sc #1}}
\newcommand{\DLMESO}{\code{dl\_meso}}
\newcommand{\DLPOLY}{\code{dl\_poly\_5}}
\newcommand{\nDPD}{\mbox{$n$\text{DPD}}}
\newcommand{\nm}{\text{nm}}
\newcommand{\kB}{k_\text{B}}
\newcommand{\kT}{\kB T}

\begin{document}

\articletype{Paper} 

\title{Construction of highly coarse-grained molecular models for aqueous electrolytes}

\author{Michael A. Seaton$^1$\orcid{0000-0002-4708-573X}, Benjamin T. Speake$^1$\orcid{0000-0002-5690-9470}, Tristan Rimmer$^1$ and Ilian T. Todorov$^{1,*}$\orcid{0000-0001-7275-1784}}

\affil{$^1$Scientific Computing, UKRI Science and Technology Facilities Council, STFC Daresbury Laboratory, Sci-Tech Daresbury, Keckwick Lane, Warrington WA4 4AD, United Kingdom}

\affil{$^*$Author to whom any correspondence should be addressed.}

\email{ilian.todorov@stfc.ac.uk}


\begin{abstract}
Molecular modelling of electrochemical systems at larger than atomic scales poses particular challenges, not least the ability to represent correct thermodynamic, hydrodynamic and dielectric properties while still being computationally viable for large scale applications. In the case of aqueous electrolytes such as those used in vanadium-based batteries with oxidation-reduction (redox) reactions, both a background medium that can respond to localised electric fields due to high-valence ions and control over ion transport properties are crucial to embed consistently in the model. To this end, we propose the use of Dissipative Particle Dynamics (DPD) combined with a newly-devised highly coarse-grained polarisable water model with pairwise interactions designed to produce realistic thermodynamic behaviour as the basis for the aqueous electrolytes found in vanadium redox flow batteries (VRFB).  We consider how the core water model can be modified to incorporate the ionic species found in VRFB catholytes and anolytes, as well as how the DPD pairwise thermostat forces can be parameterised to attain realistic diffusivities for those species.
\end{abstract}

\section{Introduction}

The ability of molecular simulation methods to assess the effectiveness of electrochemical systems depend on representation of materials and their properties at the required length and time scales. This is a particular challenge when it comes to the mesoscale, as both microscopically-informed thermodynamics and macroscopically-emergent hydrodynamics need to be accounted for when devising any such models. Systems with charged species and polarisable media (e.g. water) additionally exhibit localised variations in dielectric properties, posing further challenges when developing a mesoscopic model to accommodate these effects.

Cogent examples demonstrating these challenges include the aqueous electrolytes used in vanadium redox flow batteries (VRFB) \cite{boccardo2024,bromberger2014}. This form of battery hinges upon a reversible redox (reduction-oxidation) reaction:
\[ \ce{VO^{2+}} + \ce{V^{3+}} + \ce{H2O} \ce{<=>} \ce{VO2^+} + \ce{V^{2+}} + 2\ce{H^{+}}  \]
with the forward reaction generating charge and the backwards reaction releasing it. The catholyte includes higher charge valence vanadium ions, \ce{V^{4+}} in the form of vanadium(IV) oxide (\ce{VO^{2+}}) and \ce{V^{5+}} as vanadium(V) oxide (\ce{VO2^{+}}), as well as water (\ce{H2O}) and sulphuric acid (\ce{H2SO4}) with the latter providing protons (\ce{H^{+}}) and partially dehydrated sulphate ions (\ce{HSO4^{-}}). The anolyte has lower valence vanadium ions, \ce{V^{2+}} and \ce{V^{3+}}, along with water and sulphuric acid with the latter providing both partially and fully dehydrated (\ce{SO4^{2-}}) sulphate ions. While the two electrolytes do not come into direct contact inside a battery, they are pumped between electrodes and at least one layer of carbon felt acting as a membrane that enables protons to transfer from the catholyte to the anolyte. 

The effectiveness of this form of battery thus depends on the ability of electrolyte ions to travel between the bulk and felt surfaces, much of which is governed by diffusion through a polar solvent (water). Since the effective length scale of this process, identified as the diffusion layer thickness $\delta$, can cover several micrometres for such systems, this precludes the use of fully atomistic modelling approaches such as Molecular Dynamics (MD) and we therefore favour mesoscopic representations of the electrolytes for viable simulations. These representations require both a representation of water that can incorporate charge polarisability effects and the ability to control the diffusivities of individual ionic species. 

This study posits the use of a recently-developed highly coarse-grained polarisable water model \cite{seaton2026} as the main solvent for the catholyte and anolyte. This model exploits a pairwise particle interaction model \cite{sokhan2023} that we denote as \nDPD\ and had previously parametrised for water \cite{seaton2024} at a coarse-graining level of 5 \ce{H2O} molecules per `bead'. Attaching additional charged Drude particles to each interacting water bead using flexible harmonic bonds gives realistic dipole and quadrupole moment distributions, previously confirmed by comparing against values obtained from five-molecule clusters in an MD trajectory of atomistic (TIP3P) water. All ionic species except for protons are assumed to exist in the electrolytes as ion-water complexes are are thus treated as non-polar water beads with additional masses and charges. Control of diffusivities for all species is obtained by using and parameterising the DPD thermostat \cite{gw97}, i.e. additional pairwise dissipative and random forces that provide temperature control with automatic Galilean invariance.

\section{Model setup}

\subsection{Coarse-grained polarisable water model}

The basis of our aqueous catholyte and anolyte models for VRFB is a highly coarse-grained (CG) polarisable water model \cite{seaton2026}, where each CG entity has the equivalent mass and volume of 5 molecules of water. Fundamental to this model are \nDPD\ interactions \cite{sokhan2023} between the entities, defined by the following pairwise potential:
\begin{equation}
\label{eq:nDPDpotential}
U^{n} (r_{ij}) = \frac{A_{ij} b_{ij} r_c}{n+1} \left(1-\frac{r_{ij}}{r_c} \right)^{n+1} - \frac{A_{ij} r_c}{2} \left(1-\frac{r_{ij}}{r_c} \right)^{2}~,
\end{equation}
which gives non-zero values only when $r_{ij}<r_c$. The other parameters $A_{ij}$, $b_{ij}$ and $n$ respectively define the magnitude of attractions/repulsions, the ratio of repulsion to attraction and an exponent setting out the repulsion `steepness'. For both attraction and repulsion to exist, the value of $b_ij$ must be greater than $\frac{1}{2}\left(n+1\right)$: this provides a root $0<\sigma<r_c$ for Eq.~\ref{eq:nDPDpotential} that is derived as
\begin{equation}
\label{eq:nDPDsigma}
\frac{\sigma}{r_c} = 1 - \left(\frac{n+1}{2b_{ij}}\right)^{\frac{1}{n-1}}~.
\end{equation}
and can be used as a length scale. Setting $b_{ij}=2$ and $n=1$ reduces Eq.~\ref{eq:nDPDpotential} down to the original purely repulsion-based Groot-Warren (`standard DPD') interaction \cite{gw97}. To implement this interaction in MD and DPD codes, the derivative of Eq. \ref{eq:nDPDpotential} with respect to $r_{ij}$ can be found as the equivalent force, i.e.
\begin{equation}
\label{eq:nDPDforce}
\vec{F}^{C,n}_{ij} = A_{ij} \left[b_{ij} \left(1 - \frac{r_{ij}}{r_c} \right)^n - \left(1 - \frac{r_{ij}}{r_c} \right) \right] \frac{\vec{r}_{ij}}{r_{ij}}~,
\end{equation}
where $\vec{r}_{ij}=\vec{r}_{j}-\vec{r}_{i}$ is the vector between particles $i$ and $j$. 

Unlike the original Groot-Warren interaction from which it was derived, \nDPD\ has the capability to produce stable multiple-phase systems such as vapour-liquid coexistence below a critical point \cite{sokhan2023}. For a single particle species, the values of $A_{ij}$, $n$ and $b_{ij}$ broadly and respectively influence the critical temperature, liquid density and how the interfacial tension depends upon temperature. To that end, a set of parameters have previously been determined to represent water at a coarse-graining level of 5 molecules per bead \cite{seaton2024}, providing good matches to room temperature (298.15~K) liquid density and interfacial tension. This describes a \emph{non-polar} CG model for water that acts as the starting point for both the polarisable model that follows and the representation of most ionic species in the electrolytes.

To transform the above non-polar \nDPD\ water model into a polarisable one, we attach additional charged particles to central interacting beads \cite{chiacchiera2024,vjt18,pp14}, forming additional charge multipoles that can respond to electric fields (both external and those produced by distributed ionic species) and produce localised variations in relative permittivity. Based on our previous work creating polarisable \nDPD\ water models \cite{seaton2026}, we make use of harmonic bond potentials to attach these Drude charges to central interacting beads:
\begin{equation}
U^{b} = \frac{\kappa r_{ij}^2}{2},
\label{eq:HarmSpring}
\end{equation}
and Gaussian charge smearing \cite{wva+13,wv14} to prevent oppositely-charged particles from permanently collapsing on top of each other:  
\begin{equation}
\label{smearCoulomb}
U^{e} = \frac{\lambda_B Z_i Z_j}{r_{ij}} \left[1 - \textrm{erfc} \left(\frac{r_{ij}}{2 \sigma_G}\right) \right]~.
\end{equation}
In the above, $\kappa$ is the harmonic bond force strength, $\lambda_B=\frac{e^2}{4\pi\epsilon~k_{B}T}$ is the Bjerrum length for the background medium (using $\epsilon=\epsilon_0$ for vacuum), $Z_i=q_i/e$ is the charge valency for particle $i$ and $\sigma_G$ is a charge smearing length. Noting that Gaussian charge smearing as given in Eq.~\ref{smearCoulomb} only needs to be applied in the real space (shorter range) part of an Ewald sum, a computational advantage for its use can be found by selecting the value of the associated Ewald parameter to be $\alpha=\frac{1}{2\sigma_G}$, thus reducing all pairwise real space terms (potentials and forces) to zero and enabling all electrostatic interactions to be calculated entirely in reciprocal (Fourier) space.

\begin{figure}
    \centering
    \includegraphics[width=0.25\columnwidth]{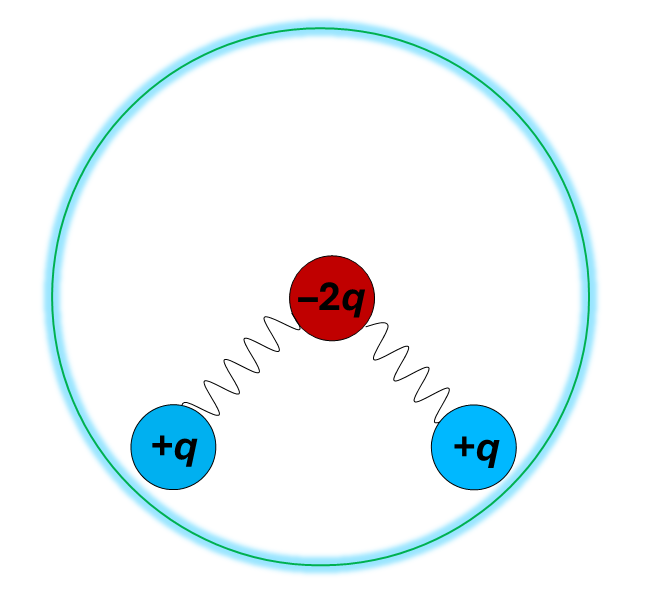}
    \caption{Chosen method to polarise \nDPD\ coarse-grained water model: a three-charge model (3CM) with two Drude charges ({\textcolor{blue}{\textbf{blue}}}, $+q$) and an additional balancing charge ($-2q$) on the interacting central bead ({\textcolor{red}{\textbf{red}}}).}
    \label{chargemodel}
\end{figure}

Our selection of a three-charge model (3CM) for our polarisable \nDPD\ water model, as illustrated in Figure \ref{chargemodel}, requires the central \nDPD-interacting bead in each entity to hold a charge of $-2e$ to balance out Drude charges of $+e$ each. While this increases the number of charge particles required for simulations compared to a two-charge model (2CM) with a neutral central bead and Drude charges of $\pm e$, it provides a greater numerical stability and the additional ability to form quadrupole moments alongside the desired dipole moments. This latter feature is likely to be important for systems involving interfaces such as vapour-liquid phase boundaries or solid boundaries (e.g. carbon felt membranes) where water molecules would arrange themselves differently than in the monophase bulk. The resulting dipole moments for each water entity, $\vec{p}_{i}~=~\sum_j q_j\vec{r}_j$, determine their relative permittivity:
\begin{equation}
\label{eq:epsilon}
\epsilon_r = 1 + \frac{4 \pi \lambda_B \langle \left(\sum_i p_i \right)^2 \rangle}{3e^2V}~,
\end{equation}
and as such, this property can be targetted by selecting charge magnitudes and the distances between charges within an entity via e.g. bond interaction parameters such as $\kappa$.  The angular brackets denote that the property inside is sampled as an ensemble average over the system's space, both configurationally and over time, from the numerical simulation.

The addition of charges required a reduction in the value of $A_{ij}$ for \nDPD\ interactions between central beads to recover the correct liquid density. The other parameters for this interaction - $b_{ij}$, $n$ and $r_c$ - remained unchanged: we previously noted \cite{seaton2026} there is potential scope to modify $b_{ij}$ and $n$ to recover other thermodynamic properties (e.g. vapour density, interfacial tension) more correctly but we choose not to do so for this study as the liquid phase is of greatest interest and importance for aqueous electrolytes.

\subsection{Ionic species}

With the notable exception of protons (\ce{H^{+}} ions), we note that all ionic species in VRFB electrolytes are commonly found to exist as ion-water complexes, i.e. each ion is surrounded by a number of water molecules as a thermodynamically stable cluster. Both vanadium ions found in the VRFB anolyte, \ce{V^{2+}} and \ce{V^{3+}}, are typically surrounded by 5 to 6 \ce{H2O} molecules \cite{krakowiak2012,Sepehr2013,Jiang2016,Vijayakumar2010,Vijayakumar2015} as primary hydration shells. The vanadium(V) oxide (\ce{VO2^{+}}) ion in VRFB catholyte typically has 3 to 4 \ce{H2O} molecules \cite{krakowiak2012,Jiang2016,Guo2024,Sepehr2013} in its primary hydration shell, while vanadium(IV) oxide (\ce{VO^{2+}}) has 5 \cite{krakowiak2012,Sepehr2013}. Meanwhile, the partially hydrated sulphate ions (\ce{HSO4^-}) are each surrounded by between 7 and 12 \ce{H2O} molecules, and fully dehydrated sulphate ions (\ce{SO4^{2-}}) are each typically surrounded by 12 water molecules \cite{Yacovitch2011,Yang2019,Kulichenko2019}. 

The fact that these ions only exist in aqueous electrolyte solutions inside ion-water complexes means that at larger than atomic scales they can be considered effectively as water beads with additional mass and charge. Beyond electrostatic interactions between ions and polarised water, each complex can be assumed to interact via \nDPD\ \emph{as though they are water} and, to simplify parameterisation, that each ion is surrounded by the same number of water molecules represented by each polarisable water entity, i.e. five. A further simplification and reduction in computational cost can be obtained from the fact that each ion-water complex has a net charge, whose effect on surrounding water molecules and other ions swamps those of the localised partial charges in its own water molecules: this means that we can use the original \emph{non-polar} \nDPD\ water model \cite{seaton2024} as the interaction basis for ion-water complexes instead of the above polarisable water model. Since both the non-polar and polarisable water models use the same values for $b_{ij}$, $n$ and $r_c$, we only additionally need to specify a value of $A_{ij}$ between central beads and ions. As a reasonable and intuitive first choice, we make use of the commonly-used Berthelot combining rule and choose the geometric mean value, i.e. $A_{ij}=\sqrt{A_{ii}A_{jj}}$. 

Protons are treated as a special case in our VRFB electrolyte models. While they do indeed form ion-water complexes, e.g. \ce{H3O+} and \ce{H5O2+}, these are not nearly as stable as those formed by heavier ions and are significantly more dynamic. As such, we treat each \ce{H^{+}} ion as a light free-moving particle that can interact with water molecules in both polarisable water entities and ion-water complexes by means of a truncated and shifted Morse potential \cite{leeMT2015}:
\begin{equation}
\label{eq:MorseProton}
U^{M}(r_{ij}) = K_{ij} \left\{ 1 - \exp \left[ -\alpha_{ij} \left(r_{ij} - r_{ij}^0\right) \right] \right\}^2 - K_{ij} \left\{1 - \exp \left[ -\alpha_{ij} \left(r_{c,ij} - r_{ij}^0\right) \right] \right\}^2~.
\end{equation}
The interaction given in Eq.~\ref{eq:MorseProton} mimics dynamic formation and dissolution of anharmonic bonds between protons and water molecules, including those surrounding vanadium and sulphate ions.

\subsection{Diffusivity control using DPD thermostat}

The interactions between particles in both the polarisable water and ionic species, both shorter-range \nDPD\ and longer-range electrostatic interactions, influence the diffusivities of those species. Since these mesoscopic-scaled interactions are somewhat simplified compared with atomistic models, it is likely that they would not \emph{on their own} correctly reproduce the diffusivities of the various species in the electrolytes. As such, we turn to pairwise dissipative and random forces defined for Dissipative Particle Dynamics (DPD) \cite{gw97} to provide additional control of this property.

The dissipative and random forces take the following forms:
\begin{equation}
\label{eq:DPDDissipative}
\vec{F}_{ij}^{D} = -\gamma_{ij} w^D \left( r_{ij} \right) \left(\hat{r}_{ij} \cdot \vec{v}_{ij}\right) \hat{r}_{ij}
\end{equation}
\begin{equation}
\label{eq:DPDRandom}
\vec{F}_{ij}^{R} = \sigma_{ij} w^R \left( r_{ij} \right) \frac{\xi_{ij}}{\sqrt{\delta t}} \hat{r}_{ij}~,
\end{equation}
where $w^D\left(r_{ij}\right)$ and $w^R\left(r_{ij}\right)$ are functions of distance between particles $i$ and $j$ ($r_{ij}$), $\vec{v}_{ij}~=~\vec{v}_j-\vec{v}_i$ is the relative velocity between the two particles, $\xi_{ij}$ is a Gaussian-distributed random number with zero mean and unit variance, $\delta t$ is the simulation timestep, and $\hat{r}_{ij}=\frac{\vec{r}_{ij}}{r_{ij}}$ is a unit vector for the particle pair. From fluctuation-dissipation theory, Eq.~\ref{eq:DPDDissipative} and \ref{eq:DPDRandom} act as a Galilean invariant, momentum-conserving thermostat provided the following conditions are met:
\begin{equation}\label{DPDcondition1}
\sigma_{ij}^2 = 2 k_BT\gamma_{ij}
\end{equation}
\begin{equation}\label{DPDcondition2}
w^D \left(r_{ij}\right) = \left[ w^R \left(r_{ij}\right) \right]^2~,
\end{equation}
noting that only the relationship between $w^D\left(r_{ij}\right)$ and $w^R\left(r_{ij}\right)$ is directly specified rather than exact functional forms, although it is evident that $w^D\left(r_{ij}\right)\geq0$ for all values of $r_{ij}$. 

It should be noted that both the values of $\gamma_{ij}$ (dissipative force parameters) and the functional forms of the switching functions influence the dynamic properties of particle species such as diffusivity and viscosity \cite{marsh1997} and, as such, we can exploit this property of the `DPD thermostat' to match the diffusivities of electrolyte species to experimental values \cite{bromberger2014}. Given the range of diffusivities available in each electrolyte is rather large, there is a risk that relying solely on one dependence in the dissipative and random forces would increase numerical instabilities during calculations. To avoid this, we use the following choice of dissipative force switching function for all distances within a cutoff $r_c$ \cite{fan2006}:
\begin{equation}
\label{eq:DPDswitch}
w^D \left( r_{ij} \right) = \left(1 - \frac{r_{ij}}{r_c} \right)^s ~~~ (r_{ij} < r_c)~,
\end{equation}
where $s$ is a parameter that provides additional control for dynamic properties, usually set as $s=2$ in many DPD simulations. In the limiting case where there are no conservative interactions (i.e. for ideal gases), the resulting self-diffusivity of a species can be determined \cite{panD2013} to be
\begin{equation}
D = \frac{3 m k_B T}{4 \pi r_c^3 \rho \gamma_{ij}} \left[\frac{1}{1+s} - \frac{2}{2+s} + \frac{1}{3+s} \right]^{-1}~,
\end{equation}
and while this relationship will not directly hold for our polarisable \nDPD\ water or ionic species, we can use its broad principles to iterate values of $\gamma_{ij}$ and $s$ for a given species to target a required diffusivity: for instance, increasing $\gamma_{ij}$ and/or decreasing $s$ should reduce diffusivity. The diffusivity of a species can be determined from an equilibrated system using temporal gradients of mean-squared displacments (MSDs):
\begin{equation}
\label{eq:MSD}
\frac{1}{N} \sum_i^N \left(\vec{r}_i (t) - \vec{r}_i (t=0) \right) = 6Dt~,
\end{equation}
providing a method to test the suitability of dissipative force parameters for each component in both electrolytes. An additional restriction we applied was to use the same dissipative force parameters for the same components in both electrolytes, i.e. water, protons and partially hydrated sulphate ions (\ce{HSO4^{-}}). While the concentrations of these components differ between the catholyte and the anolyte, these should not have a significant effect on their diffusivities. 

Given that $\gamma_{ij}$ (and $s$) can be specified for all pairs of beads, this raises the question of how to specify these parameters for non-alike pairs, e.g. between central water beads and ions. Given that the dissipative force parameters for individual species, $\gamma_{ii}$, were found to vary considerably due to different required diffusivities, masses, charge valencies and concentrations in solution, the simplest approach of using a combination rule without any modification, e.g. $\gamma_{ij}=\sqrt{\gamma_{ii}\gamma_{jj}}$, was likely to overly damp the motion of species requiring higher diffusivities. To reduce this effect and the complexity of these multi-component systems, we propose a modified Berthelot combining rule:
\begin{equation}\label{eq:mixing}
\gamma_{ij} = c \sqrt{\gamma_{ii} \gamma_{jj}}~,
\end{equation}
where $0<c\leq{1}$ is an empirically chosen parameter, one that we found could potentially be set to a constant value for all non-alike particle pairs to ensure consistency of approach.  We constrained its application to two distinct general scenarios: (i) between pairs of different aqueous species, including polarisable water beads (both central and Drude particles) and ion-water complexes but explicitly excluding protons, and (ii) between protons and any aqueous species.  Scenario (ii) includes an additional large contrast in particle masses between aqueous beads and much lighter protons: as such, this justifies the use of a different (higher) value of $c$ for these particle pairs compared with pairs of aqueous beads. To simplify the search for inter-species parameters, we set $s=2$ for all non-alike species pairs ($i\neq{j}$) but allow $s$ to vary for like-like pairs: this is particularly useful for lower diffusivity components as it enables lower $\gamma_{ii}$ values to be used, thus reducing excessive damping when the above combining rule is applied.

\section{Parameterisation and verification}

\subsection{Non-polar and polarisable water}

The parameters for the electrolytes and all of its constituent models can be expressed in terms of three selected fundamental quantities: (1) the mass of a (non-polar) water bead ($m$) based on the number of molecules per bead (coarse-graining level), (2) an energy unit based upon temperature ($\epsilon=\kT$), and (3) its radius ($\sigma$) based on the selected coarse-graining level, the values of $b_{ij}$ and $n$ for the \nDPD\ interactions, Eq. \ref{eq:nDPDsigma}, and matching up the expected bead density with the density of liquid water at the given temperature. These three quantities were subsequently used as units for DPD-based calculations and can be used to define units for other quantities such as time, pressure and electric dipoles.  All of the units in use, including their equivalent `real-life' physical values, are given in Table~\ref{tab:units}.

\begin{table}
  \centering
    \begin{tabular}{llll}
      \textrm{Physical quantity}  & DPD unit &
      Example value & unit  \\
      \hline\\[-9pt]
      mass   & $m$    & $1.49575\times10^{-22}$ & kg \\
             &        & $90.0764$ & u (Da) \\
      length & $\sigma$  & $0.623956$ & $\nm$  \\
      energy & $\epsilon = \kT$  & $4.116405\times10^{-21}$ & J \\ 
             &        & $2.47896$ & $\mathrm{kJ}\,\mathrm{mol}^{-1}$ \\
      time   & $\tau = \sigma \sqrt{m/\epsilon}$ & $3.76119$ & ps \\
      electric dipole & $e \sigma$ & $9.99688\times10^{-29}$ & $\mathrm{C}\,\mathrm{m}$ \\
                      &            & $29.9699$ & Debye \\
      electric field  & $\epsilon/(e \sigma)$
      & $4.11769 \times 10^7$ & $\mathrm{V}\,\mathrm{m}^{-1}$\\
      pressure & $\epsilon/\sigma^3$ & 16.9456 & MPa \\
      surface tension & $\epsilon/\sigma^2$ & $10.5733$
      & $\mathrm{mN}\,\mathrm{m}^{-1}$\\[-1pt]
    \end{tabular}
  \caption{DPD units and physical units for relevant physical quantities.  The values in the third column correspond to the chosen coarse-graining level (5 molecules of \ce{H2O} per water entity), room temperature ($T=298.15~\mathrm{K}$) and matching the bead density to give the equivalent value for liquid water at this temperature.\label{tab:units}}
\end{table}

Parameterisation of the original non-polar \nDPD\ water model \cite{seaton2024} hinged upon obtaining the correct liquid density for the required reduced temperature (ratio of temperature to its critical value). The original study introducing \nDPD\ \cite{sokhan2023} includes vapour-liquid phase diagram plots constructed from DPD-based simulation results for three different values of $n$ -- later confirmed via a hypernetted-chain (HNC) solver for the Ornstein-Zernicke equation with Eq.~\ref{eq:nDPDpotential} as an input \cite{seaton2024,wva+13} -- indicating how the densities of the two phases vary with reduced temperature. By reading off the plot for $n=4$ (which gave the most realistic sub-critical liquid behaviour) at the required reduced temperature and matching this to the expected liquid density at room temperature, we obtained the interaction cutoff $r_c$, which ended up being equal to $2.22381~\sigma$. We subsequently set $A_{ij}=7.56566~\epsilon\sigma^{-1}$ to give the correct critical temperature $T_c$ so that $\kT=1$ corresponds to room temperature, and selected $b_{ij}=15.0$ to match the expected interfacial tension for water. We note that this value of $b_{ij}$ not only sets the simulation length scale (particle radius) $\sigma$ via Eq.~\ref{eq:nDPDsigma}, but it also maintains thermodynamic stability and provides positive compressibility for the fluid.

Our subsequent extension of this \nDPD\ 5:1 coarse-grained water model to incorporate charge polarisability is described extensively in our previous work \cite{seaton2026} as \emph{Polar-I}, corresponding to the use of flexible harmonic bonds and, consequently, more realistic responses to polarising fields. In summary, this model applies an additional Gaussian smeared charge on the central \nDPD-interacting with a charge of $-2e$ (valency of $-2$) and attaches two (smeared) Drude charges of $+e$ each using harmonic bonds with $\kappa=47.0~\epsilon\sigma^{-2}$. The Gaussian smearing lengths for all charges ($\sigma_G=0.517929~\sigma$) were set according to previous work \cite{chiacchiera2024} in order to preserve charge screening by the water: we note that while this value can affect thermodynamic properties such as interfacial tension and calculation stability, it has no direct effect on multipole distributions. The Bjerrum length was set to the expected value for vacuum, $\lambda_B=89.8235~\sigma$, while the number of reciprocal images used for Ewald summation-based calculations was set to give a relative error in potential energy no higher than $10^{-5}$. The overall mass of each water entity is distributed among the three particles so that the central interacting bead holds two-thirds and the Drude charges each hold one-sixth to keep the centre-of-mass close to the central bead \cite{vjt18}. Simulations of periodic boxes containing the polarisable \nDPD\ water enabled the time-averaged squared dipole moments to be obtained and used in Eq.~\ref{eq:epsilon} to find its relative permittivity $\epsilon_r$. As shown in Table~\ref{tab:prop}, this value is close to the experimental value and is an expected result given the choices of $\kappa$ and charge magnitudes.

To obtain vapour-liquid coexistence in a simulation and measure the densities of the two phases and the interfacial tension, we carried out simulations in elongated boxes to encourage the water beads or entities to separate out into two distinct phases with planar interfaces between them \cite{sokhan2023}. After carrying out these calculations in either \DLMESO\ \cite{sam+13} or \DLPOLY\ \cite{dlp5} and obtaining well-equilibrated states, we transformed trajectories into time-averaged localised density plots along the longest box direction to find the vapour and liquid densities, $\rho_v$ and $\rho_l$ respectively, and calculated the interfacial tension $\gamma$ from the pressure tensors. We found that merely attaching Drude charges to each central interacting bead has the effect of changing the thermodynamic properties compared with the original non-polar \nDPD\ model. In particular, we observed a 13\% increase in liquid density for our polarisable model due to additional clustering of oppositely charged particles. To counteract this, we reduced the value of $A_{ij}$ for our polarisable water model to $4.99585~\epsilon\sigma^{-1}$, roughly two-thirds of the original value used in the non-polar \nDPD\ water model. As shown in Table~\ref{tab:prop}, this has the effect of reducing the vapour/liquid interfacial tension $\gamma$ by a similar factor. While it would be possible to correct this by modifying $b_{ij}$ and/or $n$, this would further complicate our intended use of both the non-polar and polarisable water models in the same simulation as they would require further, more extensive parameter fitting.  

Both non-polar and polarisable water models make use of the `DPD thermostat', i.e. Eq.~\ref{eq:DPDDissipative} and \ref{eq:DPDRandom}. In these cases, we selected the usual dissipative force switching function, $w^D (r_{ij})=\left(1-\frac{r_{ij}}{r_c}\right)^2$, the equivalent of setting $s=2$ in Eq.~\ref{eq:DPDswitch}. The same value for the dissipative force parameter $\gamma_{ij}=15.5~\epsilon^{1/2} m^{1/2} \sigma^{-1}$ was used for both water models for all particle pairs, while the cutoff distance for the switching function $r_c$ was set equal to the cutoff for \nDPD\ interactions ($2.22381~\sigma$) between pairs of interacting beads (e.g. central beads for polarisable water) and to $\sigma$ for all other pairs. Dissipative and random forces were still calculated between pairs of connected particles (i.e. central beads and Drude charges) to ensure correct hydrodynamic behaviour for entire polarisable water entities and adequate temperature control. We note that while our Drude charges are a computational device to incorporate charge polarisation, each polarisable water entity represents multiple water molecules and its constituent parts thus should behave consistently when considered macroscopically. 

To obtain the isothermal compressibility of the liquid non-polar and polar waters $\kappa_T$, a constant volume and temperature ($NVT$) ensemble calculation for a well-equilibrated box of each water at the liquid density $\rho_l$ was carried out to determine an ensemble-averaged pressure. This pressure was subsequently used as the target value in isotropic $NPT$ (constant pressure and temperature) ensemble calculations -- using a Langevin barostat coupled to the DPD thermostat \cite{jakobsen2005} with relaxation time $\tau_p = 5.0 \tau$ and drag coefficient $\gamma_p = 10.0 \tau^{-1}$ -- to measure the average box volume and its variance. The isothermal compressibilities for both water models shown in Table~\ref{tab:prop} were then obtained using:
\begin{equation}
\kappa_T = \frac{\langle V^2 \rangle - \langle V \rangle^2}{\langle \kT \rangle \langle V \rangle}~.
\end{equation}
These values are estimated to be around 20 to 30 times higher than the experimental value for water, with the non-polar model producing a larger over-estimation of this property. 

To obtain the hydrodynamic properties given in Table~\ref{tab:prop}, we carried out long simulations of each type of water after equilibration in both \DLMESO\ and \DLPOLY. The self-diffusivity $D$ was obtained from mean-squared displacements (MSDs) using Eq.~\ref{eq:MSD} for simulations in periodic boxes, while the dynamic viscosity $\mu$ was obtained by applying Lees-Edwards boundary conditions \cite{lees1972} to invoke linear shear with a range of shear rates (velocity gradients) and to measure the resulting constant shear stresses. In addition to confirming that the polarisable water model produced a constant viscosity and is therefore a Newtonian fluid, we found that transforming the \nDPD\ water model to make it polarisable by attaching charges had no significant effect on either the self-diffusivity or the dynamic viscosity. Noting that both the conservative interactions and DPD thermostatting forces contribute to the diffusivity and viscosity for a modelled fluid \cite{espanol1995}, the fact that these properties are nearly constant suggests we have successfully maintained the Hamiltonian of the original \nDPD\ water model after polarising it, and thus ensuring that the embeddiment of new physics in our polar model does not alter the dynamical behaviour of the original apolar model.

\begin{table*}
  \centering
  \begin{tabular}{l|c|c|c}
    \textrm{Property} & Non-polar & Polar & Experimental \\
    \hline\\[-9pt]
    $\rho_l$/kg~m$^{-3}$ & 1007.0 & 999.3 & 997.1 \\
    $\rho_v$/kg~m$^{-3}$ & 2.59 & 6.56 & 0.023 \\
    $\gamma$/$10^{-3}$~N~m$^{-1}$ & 73.0 & 48.5 & 72.0 \\
    $D$/$10^{-9}$~m$^2$~s$^{-1}$ & $2.50 \pm 0.48$ & $2.47 \pm 0.47$ & 2.43 \\
    $\mu$/$10^{-4}$~Pa~s & $10.33 \pm 0.21$ & $10.71 \pm 0.27$ & 8.891 \\
    $\kappa_T$/$10^{-10}$~Pa$^{-1}$ & $124.3$ & $90.27$ & $4.60$ \\
    $\epsilon_r$ & 0.0 & $78.1 \pm 1.4$ & 78.4 \\
  \end{tabular}
  \caption{Properties obtained from \nDPD\ non-polar and polarisable water models compared with experimentally determined values at $298.15~\mathrm{K}$: liquid density $\rho_l$, vapour density $\rho_v$, interfacial tension $\gamma$, self-diffusivity $D$, dynamic viscosity $\mu$, isothermal compressibility $\kappa_T$ and relative permittivity $\epsilon_r$.\label{tab:prop}}
\end{table*}

\subsection{Electrolytes}

\subsubsection*{Setting up solutions:} We start our study of the VRFB catholyte and anolyte by considering their compositions \cite{bromberger2014}. We started by specifying an initial volume for a simulation box of each electrolyte, $V=32^3~\sigma^3$: if only water were included in this box, the target liquid density $\rho_l$ would require 53,066 polarisable water entities. The box volume and the concentrations provide the numbers of ions to include in the box for each electrolyte, ultimately giving the mole fractions for each component. These mole fractions were then used to determine the numbers of beads or entities for each electrolyte simulation as given in Table~\ref{tab:conc}, noting that (i) the number of water particles is always a fifth of the number of molecules due to the selected coarse-graining degree, and (ii) each ion other than a proton also includes five water molecules, thus reducing the number of polarisable water entities needed for each electrolyte. 

\begin{table*}
  \centering
  \begin{tabular}{l|c|c|c|c|c|c}
    \multirow{2}{*}{\textrm{Component}} & \multicolumn{2}{c|}{Concentration/mol~L$^{-1}$} & \multicolumn{2}{c|}{Mole fractions} & \multicolumn{2}{c}{Number of beads/entities} \\
    & Catholyte & Anolyte & Catholyte & Anolyte & Catholyte & Anolyte\\
    \hline\\[-9pt]
    \ce{V^{5+}} (\ce{VO2^{+}}) & 0.8 & 0.0 & 0.01279 & 0.00000 & 3,351 & 0 \\
    \ce{V^{4+}} (\ce{VO^{2+}}) & 0.8 & 0.0 & 0.01279 & 0.00000 & 3,351 & 0 \\
    \ce{V^{3+}} & 0.0 & 0.8 & 0.00000 & 0.01304 & 0 & 3,701 \\
    \ce{V^{2+}} & 0.0 & 0.8 & 0.00000 & 0.01304 & 0 & 3,701 \\
    \ce{HSO4^{-}} & 4.0 & 3.6 & 0.06395 & 0.05868 & 16,753 & 16,654 \\
    \ce{SO4^{2-}} & 0.0 & 0.4 & 0.00000 & 0.00652 & 0 & 1,851 \\
    \ce{H^{+}} & 1.6 & 0.4 & 0.02558 & 0.00652 & 6,700 & 1,851 \\
    Water & 55.35 & 55.35 & 0.88489 & 0.90220 & 22,911 & 25,308
  \end{tabular}
  \caption{Concentrations and mole fractions of ionic components in vanadium redox flow battery catholyte and anolyte \cite{bromberger2014}, along with the numbers of beads or polarisable water entities used in \nDPD\ simulations.  Note that, as described in the text, the  number of (polarisabole) water beads in each electrolyte is partially reduced with respect to their mole fraction due to water beads placed on the salt ions to embed then in a water complex.\label{tab:conc}}
\end{table*}

\subsubsection*{Extending interactions:} Charge-based interactions between the various ions and the polarisable water entities use the same basis as for the polarisable water alone: Gaussian charge smearing (Eq.~\ref{smearCoulomb}) with a length scale of $\sigma_G=0.517929~\sigma$, a Bjerrum length for vacuum of $\lambda_B=89.8235~\sigma$ and reciprocal space vectors corresponding to a maximum energy truncation error of $10^{-5}$. The non-charged (short-range) interactions between the newly-introduced ion-water complexes are based on the original \nDPD\ non-polar water model, i.e. Eq.~\ref{eq:nDPDpotential} and \ref{eq:nDPDforce} with $A_{ij}=7.56566~\epsilon\sigma^{-1}$, $b_{ij} = 15.0$, $n=4$ and $r_c=2.22381~\sigma$. Interactions between the ion-water complexes and the central polarisable water beads use the same $b_{ij}$, $n$ and $r_c$ values with the geometric mean value of $A_{ij}$ between the two models, i.e. $A_{ij}=6.15791~\epsilon\sigma^{-1}$. No interactions other than smeared-charge electrostatics are used between the polarisable water Drude charges and all ions, i.e. protons and ion-water complexes, thus leaving water beads' polarisability response purely a function of carge-charge interactions across the system's environment. Interactions between protons and all interacting water beads (both polarisable water central beads and ion-water complexes) are given by Eq.~\ref{eq:MorseProton} using $K_{ij}=15.4461~\epsilon$, $\alpha_{ij}=1.0$, $r_{0,ij}=0.0$ and $r_{c,ij}=0.621515~\sigma$ \cite{leeMT2015}.

\subsubsection*{Tackling pressure:} The inclusion of ions in the VRFB electrolyte systems and additional interactions -- based on both \nDPD\ and electrostatics -- extend the original polar water model and affect the resulting equation of state, leading to changes in thermodynamic properties. While the mass densities of the two electrolytes based on the compositions in Table~\ref{tab:conc} and the initial volume are 1316.7 kg~m$^{-3}$ and 1415.8 kg~m$^{-3}$ for the catholyte and anolyte respectively -- reasonably close to the expected experimentally-determined densities of 1341.5 kg~m$^{-3}$ and 1372.4 kg~m$^{-3}$ \cite{bromberger2014} -- these lead to unphysical negative pressures in constant volume ensemble ($NVT$) simulations.

To circumvent this problem, we re-used the pressure obtained from a pure, well-equilibrated liquid polarisable water system using an $NVT$ ensemble that we applied in our previous isothermal compressibility calculations. The measured pressure -- $0.054330~\epsilon\sigma^{-3}$, equivalent to 920.7~kPa -- is significantly higher than standard atmospheric pressure but directly results from the interactions used in the polarisable water model (providing for dielectric response) and can be considered a representative condition for the electrolytes. We note at this point that while the DPD thermostatting forces can contribute towards the system virial and pressure, these contributions should (and do) become negligibly small for well-equilibrated periodic systems. This pressure was used as the target pressure for isotropic $NPT$ (constant pressure and temperature) ensemble calculations to determine representative volumes for each electrolyte: we employed the Nos\'{e}-Hoover thermostat and Berendsen barostat with respective coupling parameters of $\tau_t=0.1~\tau$ and $\tau_p=0.5~\tau$. While the resulting mean box volumes -- $27.361^3~\sigma^3$ for the catholyte, $27.883^3~\sigma^3$ for the anolyte -- give densities (2106.4~kg~m$^{-3}$ and 2140.0~kg~m$^{-3}$) that are significantly higher than experimental values, the difference in densities between the two electrolytes is reasonably close to the expected value, and the anolyte is still predicted to be more dense than the catholyte. We note that this overestimation in electrolyte densities is predominately due to the overly high compressibilities produced by both the non-polar and polarisable water models.

\begin{table*}
  \centering
  \begin{tabular}{l|c|c|c|c|c}
    Component &  \multicolumn{5}{c}{Diffusivities, $D$/$10^{-9}$~m$^{2}$~s$^{-1}$} \\
    & Exp. & Catholyte $NPT$& Anolyte $NPT$& Catholyte $NVT$& Anolyte $NVT$\\
    \hline\\[-9pt]
    Water (\ce{5H2O}) & $2.43$ & $26.08$ & $22.71$ & $26.11$ & $22.68$ \\
    \ce{V^{5+}} (\ce{VO2^{+}}) + \ce{5H2O} & $0.39$ & $20.53$ & - & $21.19$ & - \\
    \ce{V^{4+}} (\ce{VO^{2+}}) + \ce{5H2O} & $0.39$ & $15.62$ & - & $15.42$ & - \\
    \ce{V^{3+} + 5H2O} & $0.24$ & - & $8.40$ & - & $8.41$ \\
    \ce{V^{2+} + 5H2O} & $0.24$ & - & $10.81$ & - & $10.36$ \\
    \ce{HSO4^{-} + 5H2O} & $1.33$ & $24.61$ & $17.66$ & $24.57$ & $17.33$ \\
    \ce{SO4^{2-} + 5H2O} & $1.07$ & - & $12.62$ & - & $12.10$ \\
    \ce{H^{+}} & $9.31$ & $136.5$ & $107.1$ & $131.9$ & $113.2$ 
  \end{tabular}
  \caption{Diffusivities of VRFB electrolyte components obtained using the Berendsen barostat with Nos\'{e}-Hoover thermostat ($NPT$ ensemble) and the Nos\'{e}-Hoover thermostat alone ($NVT$ ensemble), compared with experimentally etermined values \cite{bromberger2014}.  Note that both sets of results use ensembles that make no attempt to control diffusivity, i.e. they do not include the DPD thermostat, in order to obtain the inherent kinetic behaviour of the CG electrolytes.  The total smapling error of non-DPD diffusivities is below 6\%.}\label{tab:npt}
\end{table*}

\subsubsection*{Diffusivity matters:} Our coarse-grained (CG) molecular dynamics (MD) calculations using $NPT$ (Berendsen barostat coupled with Nos\'{e}-Hoover thermostat) and $NVT$ (Nos\'{e}-Hoover thermostat) integrators enabled us to determine how the conservative interactions in our electrolyte models could affect the diffusivities of their components without involving the DPD thermostat (Eq.~\ref{eq:DPDDissipative} and \ref{eq:DPDRandom}). Table~\ref{tab:npt} indicates the diffusivities for each component in both electrolytes as measured during these calculations using mean-squared displacements (MSDs).  Compared with experimentally determined values \cite{bromberger2014}, we note that the measured diffusivity of each component, while the same within statistical error for both $NPT$ and $NVT$ integrators -- the latter using averaged volumes and a higher coupling time, $\tau_t = 5.0 \tau$ -- is at least an order of magnitude too high. These large differences indicate that the CG representation of the electrolytes on its own cannot reproduce correct kinetic properties, thus demonstrating the importance and utility of the DPD thermostat to adjust these properties to better match experimental observation and also noting its previous success in doing so for both non-polar and polarisable water models. It is worth noting that running the electrolytes in the microcanonical ensemble ($NVE$) leads to a gradual temperature rise, confirming the need for temperature control to preserve the balance between structure and thermodynamics in the space of CG fluids.

\begin{table*}
  \centering
  \footnotesize
  \setlength{\tabcolsep}{4pt}
  \begin{tabular}{l|c|c|c|c|c|c|c|c|c}
    Component & $m_i$ & $Z_i$ & $\gamma_{ii}$ /  & $s$ &\multicolumn{2}{c|}{Bead fractions} & \multicolumn{3}{c}{Diffusivities, $D$/$10^{-9}$~m$^{2}$~s$^{-1}$} \\
    & & & $\epsilon^{1/2} m^{1/2} \sigma^{-1}$ & & Catholyte & Anolyte & Exp. & Catholyte & Anolyte \\
    \hline\\[-9pt]
    Water (central beads) & $0.666$ & $-2$ & $9$ & $2$ & $0.23169$ & $0.24409$ & \multirow{2}{*}{$2.43$} & \multirow{2}{*}{$2.46 \pm 0.30$} & \multirow{2}{*}{$2.24 \pm 0.27$} \\
    Water (Drude charges) & $0.167$ & $+1$ & $9$ & $2$ & $0.46337$ & $0.48818$ & & & \\
    \ce{V^{5+}} (\ce{VO2^{+}}) + \ce{5H2O} & $1.9208$ & $+1$ & $160$ & $0.005$ & $0.03389$ & - & $0.39$ & $0.362 \pm 0.173$ & - \\
    \ce{V^{4+}} (\ce{VO^{2+}}) + \ce{5H2O} & $1.7432$ & $+2$ & $130$ & $0.005$ & $0.03389$ & - & $0.39$ & $0.385 \pm 0.166$ & - \\
    \ce{V^{3+} + 5H2O} & $1.5655$ & $+3$ & $135$ & $0.005$ & - & $0.03570$ & $0.24$ & - & $ 0.235 \pm 0.115 $ \\
    \ce{V^{2+} + 5H2O} & $1.5655$ & $+2$ & $200$ & $0.005$ & - & $0.03570$ & $0.24$ & - & $0.237 \pm 0.128$ \\
    \ce{HSO4^{-} + 5H2O} & $2.0776$ & $-1$ & $48$ & $2$ & $0.16941$ & $0.16063$ & $1.33$ & $1.37 \pm 0.20$ & $ 1.31 \pm 0.18$ \\
    \ce{SO4^{2-} + 5H2O} & $2.0664$ & $-2$ & $25$ & $0.75$ & - & $0.01785$ & $1.07$ & - & $1.05 \pm 0.50$ \\
    \ce{H^{+}} & $0.01119$ & $+1$ & $32$ & $2$ & $0.06775$ & $0.01785$ & $9.31$ & $9.37 \pm 1.77$ & $8.57 \pm 3.06$
  \end{tabular}
  \caption{Diffusivities of VRFB electrolyte components obtained using the DPD thermostat, at state volume and pressure of the pure liquid polarisable water system, compared with experimentally-determined values \cite{bromberger2014}.  Note that the water bead fractions are smaller than the corresponding elemental mole fractions due to corrections to account for water associated with salt ions in the electrolytes.}\label{tab:diffusivity}
\end{table*}

\begin{figure}
    \centering
    \includegraphics[width=0.8\columnwidth]{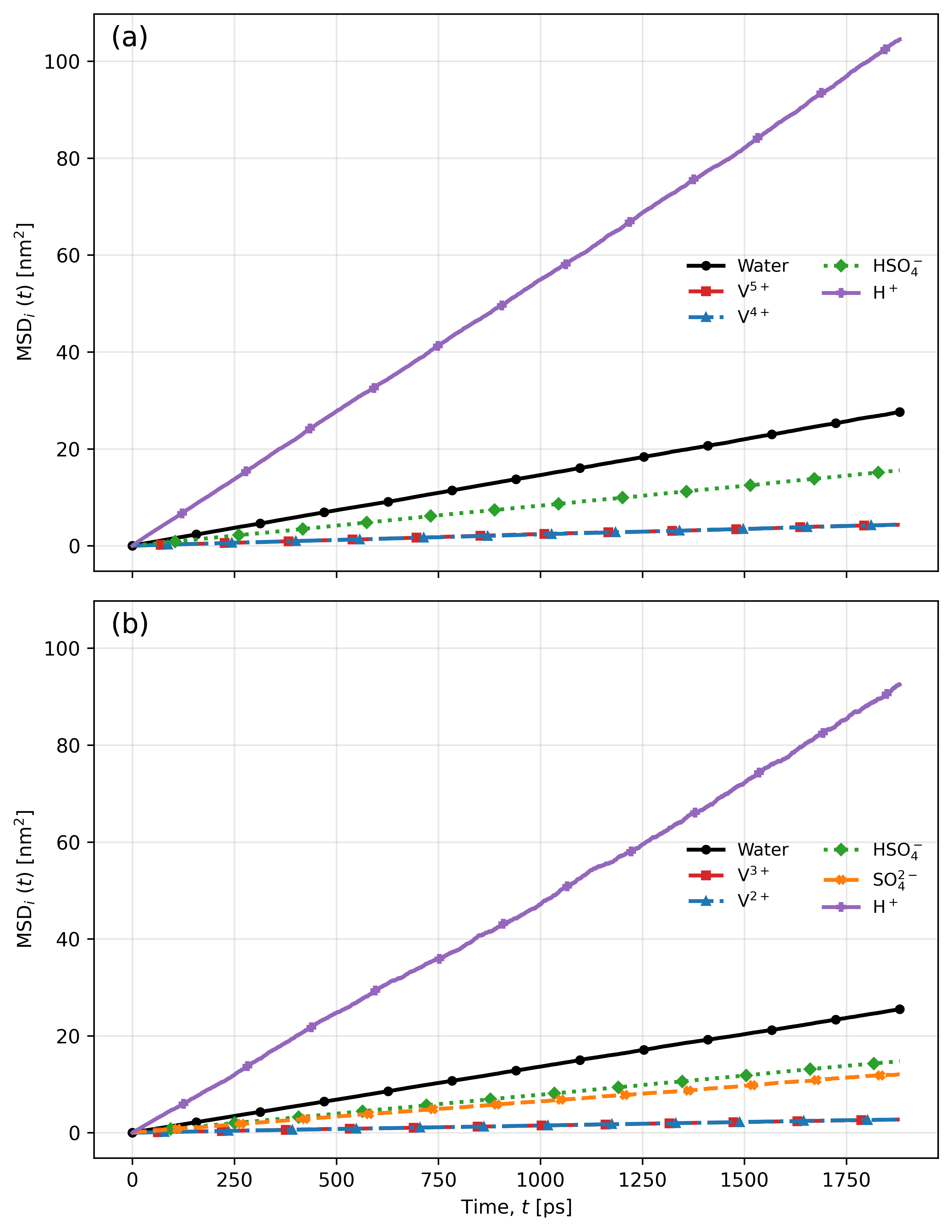}
    \caption{Mean-squared displacements of bead/entity species in coarse-grained VRFB electrolytes as functions of time: (a) catholyte, (b) anolyte.}
    \label{msd}
\end{figure}

To bring the diffusivities of the components in each electrolyte down towards their experimental values, we swapped out the Nos\'{e}-Hoover thermostat with DPD, i.e. the pairwise dissipative and random forces given in Eq.~\ref{eq:DPDDissipative} and \ref{eq:DPDRandom} respectively, using a second-order Shardlow split integrator implementation \cite{shardlow2003}. Unlike the Nos\'{e}-Hoover coupling parameter $\tau_t$, the dissipative force parameters $\gamma_{ij}$ and the functional form of these forces strongly influence the kinetic properties of each and every component in a system. Our choice of Eq.~\ref{eq:DPDswitch} for the switching function used in Eq.~\ref{eq:DPDDissipative} (with the cutoff distance $r_c$ and power index $s$ as pairwise parameters), along with the conditions given in Eq.~\ref{DPDcondition1} and \ref{DPDcondition2}, enable us to fully specify the DPD thermostat for each pair of bead species.

\subsubsection*{Constraining the DPD parameter space:} We additionally made use of a number of assumptions to simplify our search for parameters:
\begin{itemize}
\item The polarisable water is treated in a similar manner as before: a single $\gamma_{ij}$ value for all pairs of central beads and Drude charges, $s=2$ for all pairs and $r_c$ either set to the value used for \nDPD\ interactions ($2.22381\sigma$) for pairs of central beads or to $\sigma$ for all other water bead pairs (i.e. those involving Drude charges);
\item Dissipative force parameters between two non-alike species $\gamma_{ij}$ are derived from values for each individual species, $\gamma_{ii}$ and $\gamma_{jj}$, by using the combining rule in Eq.~\ref{eq:mixing} with (ideally) a constant value for its scaling factor $c$ among all non-alike species pairs;
\item The value of $s$ for non-alike species pairs was fixed to its default value of 2, while this value could vary for pairs of the same bead type to prevent $\gamma_{ii}$ from needing to be overly high (particularly for low diffusivity species such as vanadium);
\item The switching function cutoff distance $r_c$ for pairs of beads that interact directly with \nDPD\ (Eq.~\ref{eq:nDPDpotential}) or Morse (Eq.~\ref{eq:MorseProton}) potentials is the same cutoff distance as for that interaction, i.e. $2.22381~\sigma$ for \nDPD-interacting beads (all water and ions in water complexes), and $0.62151~\sigma$ for Morse (all proton-water pairs);
\item The switching function cutoff distance for pairs of protons that only interact electrostatically is set to $r_c=\sigma$ (in a similar spirit to that of Drude charges), while the cutoff distances for pairs of ion-water beads and Drude charges are set to the arithmetic mean of the \nDPD\ cutoff and $\sigma$, i.e. $r_c=1.61190~\sigma$;
\item No variations in $\gamma_{ii}$, $s$ or $r_c$ were assumed for different concentrations or mole fractions of species found in both electrolytes, i.e. water, partially dehydrated sulphate (\ce{HSO4^{-}}) or protons.
\end{itemize}
These assumptions reduced the parameter space search down to finding $\gamma_{ii}$ and $s$ values for all same-species pairs (keeping these constant for all polarisable water beads), as well as appropriate values for $c$ in the combining rule. We discovered during the parameter search that setting $c=0.1$ for all particle pairs other than those involving protons was sufficient to prevent lower diffusivity species from significantly reducing the mobility of higher diffusion species while still ensuring good temperature control.  This was particularly important to ensure the slowest-moving vanadium ions, which required the largest $\gamma_{ii}$ values, did not overly damp the motion of water in each electrolyte. The low mass of protons meant that it was advantageous to damp their motion more than other species: as such, we discovered that setting $c=1$ for species pairs involving protons, i.e. resorting to the standard Berthelot combining rule, worked sufficiently well.

\subsubsection*{Search procedure and final verification:} To carry out our search for $\gamma_{ii}$ and $s$ values, we exploited \DLPOLY's on-the-fly sampling of mean-squared displacements (MSDs) and reporting of resulting diffusivities for each species at the end of a calculation after equilibration. The reported diffusivities and how they compared with the experimentally measured (target) values informed new $\gamma_{ii}$ and $s$ values to try, which together with the combining rule (Eq.~\ref{eq:mixing}) and the chosen values of $c$ enabled us to update the remaining dissipative force parameters before launching a new calculation. These calculations used an implementation of second-order Shardlow splitting to integrate the DPD thermostat forces \cite{shardlow2003} (Eq.~\ref{eq:DPDDissipative} and \ref{eq:DPDRandom}) with a timestep size of $\delta t = 0.001\tau$ to keep force integration errors low for MSD sampling over 100,000 timesteps after 100,000 timesteps of equilibration. We noted that changing from the Nos\'{e}-Hoover $NVT$ ensemble to the DPD thermostat increased the time taken by \DLPOLY\ to complete one timestep on 64 processor cores of SCARF from 26.97 ms to 92.49 ms for the catholyte and 51.60 ms to 232.5 ms for the anolyte. This approximate quadrupling of computational time can be attributed to a change from per-particle to pairwise thermostatting, but we argue this additional cost is offset by the DPD thermostat's ability to actively control kinetic properties.

Once we finally obtained the $\gamma_{ii}$ and $s$ values given in Table~\ref{tab:diffusivity}, additional calculations were carried out using \DLMESO\ to obtain per-species MSD values as functions of time (Figure \ref{msd}).  These MSD values were processed more thoroughly to verify the diffusivities for each species in the electrolytes and obtain error bars for their values, as reported in Table~\ref{tab:diffusivity}. The same form of DPD thermostat force integration and timestep size were used for these calculations, with MSDs collected every 20 timesteps over 500,000 timesteps after 500,000 timesteps of equilibration. To verify that no aggregation was present in the bulk electrolytes, we also calculated time-averaged localised densities of all ionic bead species along each Cartesian direction: while these are not shown here, we observed that each component's local density remained constant throughout the box volume. 

The mean measured diffusivity values for all species are within $\pm 8\%$ of their intended experimental values. We note the higher densities for the electrolytes require lower dissipative force parameters for the polarisable water beads in order to obtain the correct self-diffusivity: $\gamma_{ij} = 9 \epsilon^{1/2} m^{1/2} \sigma^{-1}$ compared with $\gamma_{ij} = 15.5 \epsilon^{1/2} m^{1/2} \sigma^{-1}$ for pure polarisable water. Comparisons of the ion diffusivities between Tables \ref{tab:npt} and \ref{tab:diffusivity} indicate that higher masses and larger absolute charge valencies typically lead to lower diffusivities even without using the DPD thermostat. While the dissipative force parameters $\gamma_{ii}$ for all ionic species are still higher than those needed for water, and the vanadium ions all require lower $s$ values to obtain the lowest available diffusivities in the electrolytes, there are still negative correlations between $\gamma_{ii}$ and valency $Z_i$ or bead mass $m_i$. For species found in both electrolytes, there is barely any correlation between measured diffusivities and their compositions (shown in Table~\ref{tab:diffusivity} as fractions of beads), although the bead compositions more strongly influence the error bars obtained for those diffusivities. This is best illustrated in Figure~\ref{msd} by the example of proton (\ce{H^{+}}) beads: the smaller bead fraction in the anolyte produces a less clearly linear function of MSD with respect to time compared with the catholyte, leading to greater uncertainty in its gradient and $D$ value. 

\begin{figure}
    \centering
    \includegraphics[width=0.8\columnwidth]{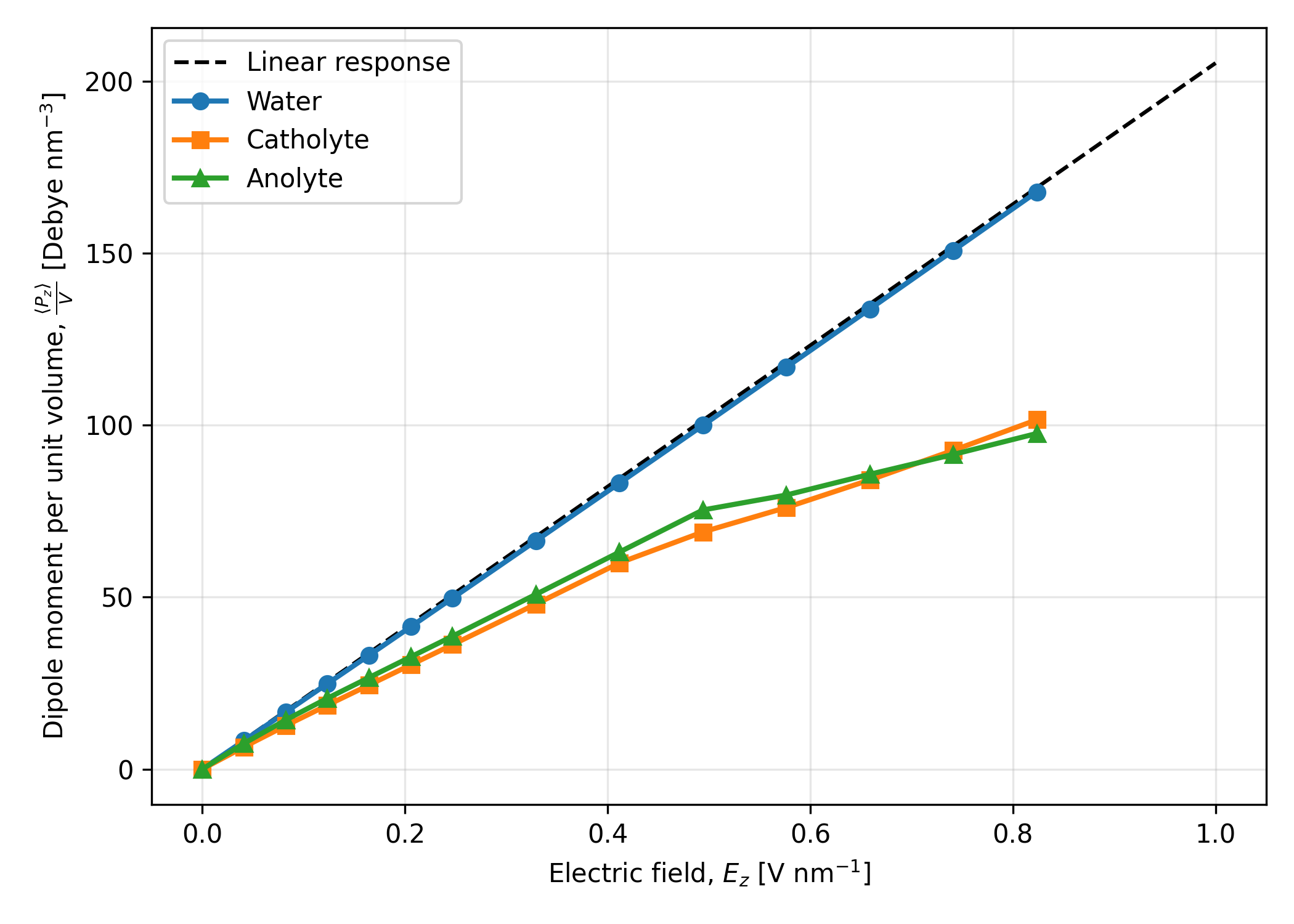}
    \caption{Induced polarisability response (overall dipole moment divided by volume) as a function of constant electric field for polarisable water ({\textcolor{blue}{\textbf{blue}}}), catholyte ({\textcolor{orange}{\textbf{orange}}}) and anolyte ({\textcolor{green}{\textbf{green}}}), compared with continuum electrostatic theory for water (dashed \textbf{black} line).}
    \label{elecfieldpolar}
\end{figure}

\subsubsection*{External electric field application:}
To explore how our electrolyte models fared with respect to the background solvent (water) model, we carried out DPD simulations with constant electric fields of different strengths along a selected direction, as we had previously carried out for the original polarisable water models \cite{seaton2026}.  This was achieved by including an additional force on each charged particle equal to $q_i~\vec{E}$, where $\vec{E}$ is a constant electric field. System properties are calculated as ensemble averages, annotated with angled brackets as $\langle property\rangle$, using a second-order Shardlow splitting integration scheme for the DPD thermostat \cite{shardlow2003} with a timestep $\Delta t=0.00125\tau$ -- sufficiently small to not limit the discretisation errors that could arise from the fast dynamics of the lightest beads, \ce{H^{+}} ions, in the electrolytes -- and collected over an interval of 200,000 timesteps after 5,000 timesteps of equilibration.

Figure \ref{elecfieldpolar} compares and contrasts polarisability reactions, given as the induced dipole moment for each system's dipoles (i.e. polarisable water entities) divided by its volume, 
${\langle P_z\rangle}{/V}={{\langle\sum_i}{q_i}{r_{i,z}}\rangle}{/V}$ between the two electrolyte models and that of pure polarisable water, when subjected to an external electric field, $\vec{E}=\left(0,0,E_z\right)$. Classical continuum electrostatic theory predicts a linear response with respect to the induced electric field:
\begin{equation}
\frac{\langle P_z \rangle}{V}=\epsilon_0\left(\epsilon_r-1\right)E_z=\frac{e^2\left(\epsilon_r-1\right)}{4\pi\lambda_B\kT}E_z~,
\end{equation} 
and the plot in Figure~\ref{elecfieldpolar} for pure water confirms this trend, predicting a relative permittivity of $77.8$, slightly lower than the value obtained without an applied electric field and Eq.~\ref{eq:epsilon}, $\epsilon_r = 78.1~\pm~1.4$ as given in Table \ref{tab:prop}. For the two electrolytes, the inclusion of single ions in place of the polarisable water leads to reductions in relative permittivity: looking at smaller electric fields where linear behaviour is still observed, we estimate permittivities for the catholyte and anolyte of $\epsilon_{r}^{\text{cath}} = 54.9~\pm~2.2$ and $\epsilon_{r}^{\text{anol}} = 56.8~\pm~3.5$ respectively.

These reductions in relative permittivity may initially seem somewhat counter-intuitive, especially given that the relative permittivity of anhydrous sulphuric acid is even larger than that of water, the inclusion of ion-water complexes in water generally promotes the stability of hydrogen bonding networks, and its inclusion in the VRFB electrolytes is most abundant with respect to that of other ions, particularly \ce{V} and \ce{H^{+}}.  We can, however, explain this phenomenon by means of a simple permittivity decrement model.  Assuming that the permittivity of water is a function of its hydrogen bonding connectivity as facilitated by its polarisability (existence of internal dipoles), then adding an ion with a charge of $q_i$ disrupts this connectivity and reduces the permittivity of the background solvent due to formation of a hydration shell of strongly bound water molecules -- losing dipole-reactive water molecules from the solvent medium -- on the one hand, and on the other to the charge density of its own electric field further disrupting the polarisability of the free water molecules near its hydration shell.  Therefore, the sum of the reciprocal charge densities of all the ions in a water solution could serve as a comparative measure of the permittivity decrement of the water permittivity based on their concentrations.  For a given ion $i$, any one of its ionic radius $R_i$, volume ($V_i \approx R_i^3$) or the average number of water molecules surrounding it $w_i$ can be used to determine the size of its hydration shell, which we can denote generally as $\phi_i$. We can then obtain a permittivity decrement number, $n_{pd}$, using any of the three hydration shell measures, by summing up products of these with the ion's mole fractions, $x_i$, divided by their absolute net charges, $|q_i|$, i.e.:
\begin{equation}
\label{eq:pdm}
n_{pd}=\sum_{i\in\textrm{ions}}~x_i\phi_i/|q_i|~. 
\end{equation}
It is worth noting that the \ce{H^{+}} ion has a negligible ionic radius and no hydration shell, and is thus excluded from the above sum.

\begin{figure}
    \centering
    \includegraphics[width=0.8\columnwidth]{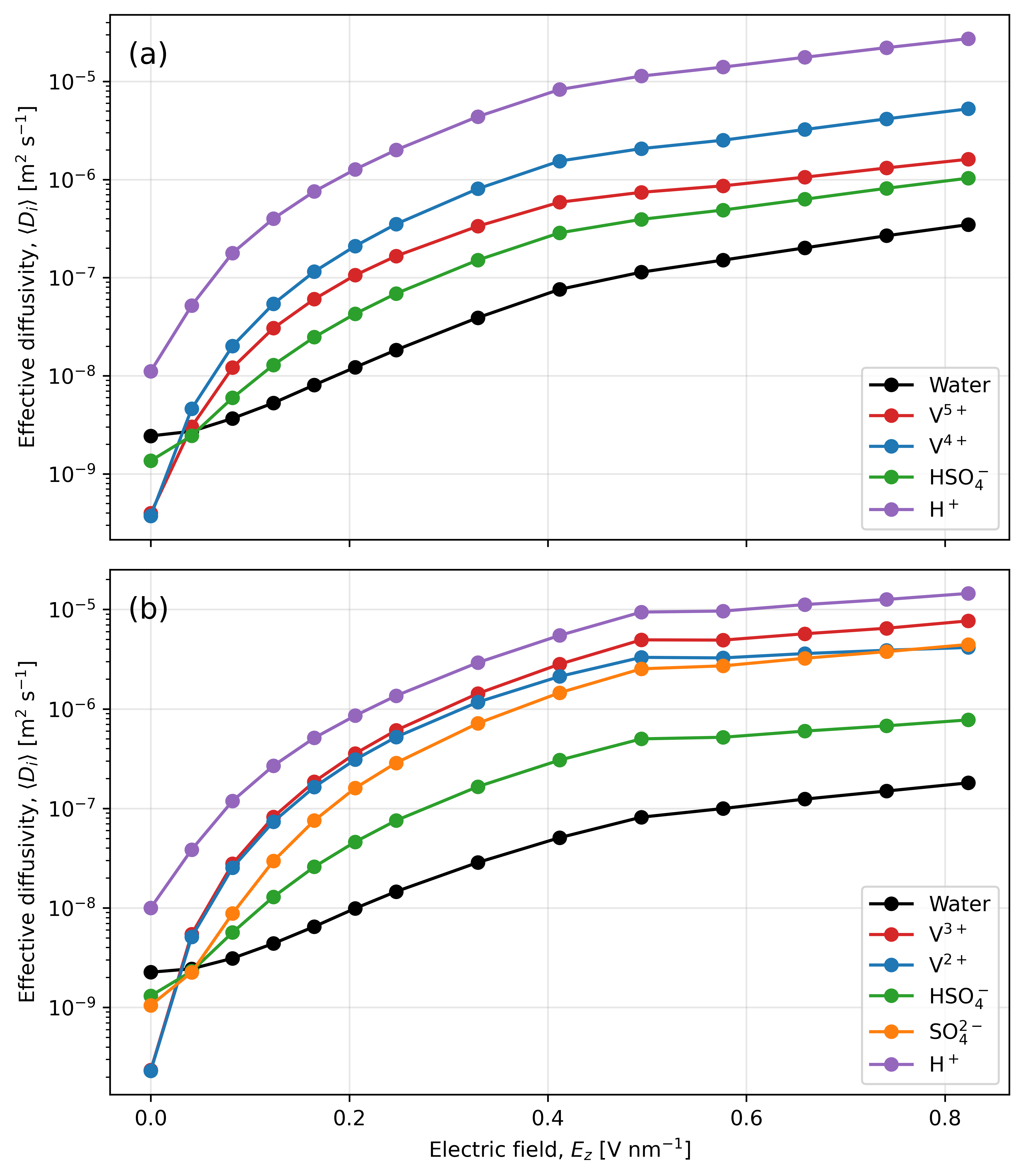}
    \caption{Effective diffusivities $\langle D_i\rangle$ of components in (a) catholyte and (b) anolyte when subjected to constant electric fields (logarithmic $y$-axes for $\langle D_i\rangle$ values).}
    \label{elecfielddiffusion}
\end{figure}

\begin{figure}
    \centering
    \includegraphics[width=0.8\columnwidth]{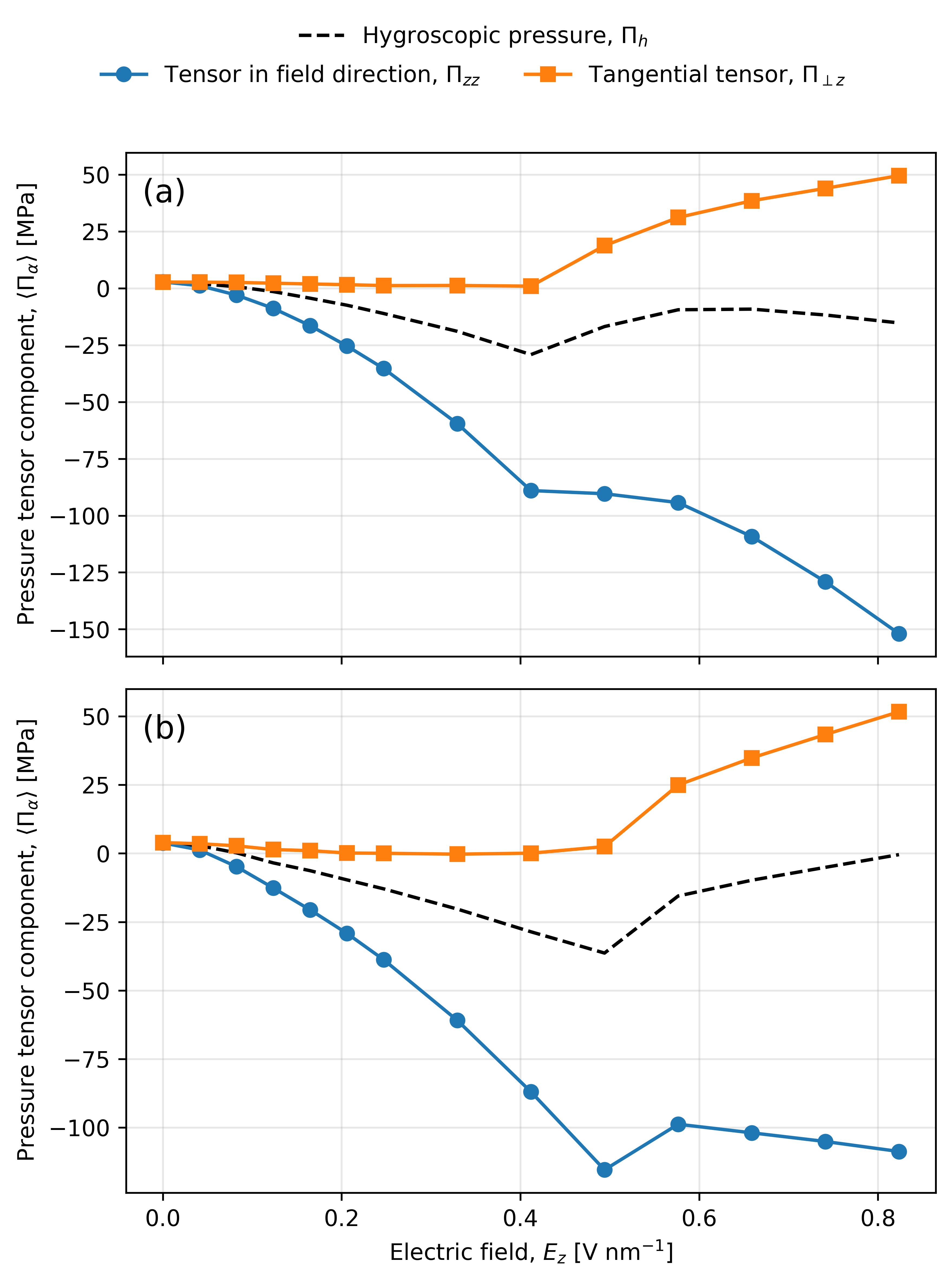}
    \caption{Induced pressure tensor components -- hygroscopic - $\langle\Pi_h\rangle$ dashed \textbf{black} line, in the direction of electric field - $\langle\Pi_z\rangle$ solid {\textcolor{blue}{\textbf{blue}}} line, and orthogonally - $\langle \Pi_{\perp z}\rangle$ solid {\textcolor{orange}{\textbf{orange}}} line -- as functions of constant electric field for (a) catholyte and (b) anolyte.}
    \label{elecfieldpressureelectrolyte}
\end{figure}

Given the above definition and the experimental data our VRFB electrolyte models are based upon, the permittivity decrement number of the catholyte is larger than that of the anolyte.  This is also confirmed by our simulations via Figure~\ref{elecfieldpolar}, where one can visually compare the linear trends of $\langle P_z\rangle$ on $E_z$ for both VRFB electrolyte models.  However, while this linear dependence holds extremely well to high electric field strengths for the polar water model, this is not the case for the VRFB electrolyte models.  Both models show deviations from their own linear response at critical field strengths of around $10 \epsilon/(e\sigma)$ ($E_{c}^{\text{cath}}\approx0.4118$~V~nm$^{-1}$) for the catholyte, and at around $11.5 \epsilon/(e\sigma)$ ($E_{c}^{\text{anol}}\approx0.4735$~V~nm$^{-1}$) for the anolyte.  We correlate these changes at the same $E_{c}$ values with changes of induced flow regimes in each electrolyte, affecting dramatically the trends of both (i) the effective diffusivities for each species, $\langle D_{i}\rangle$, given in Figure~\ref{elecfielddiffusion}, and (ii) the components of ensemble-averaged pressure components, hygroscopically $\langle\Pi_h\rangle=\langle\Pi_{xx}+\Pi_{yy}+\Pi_{zz}\rangle{/3}$, in the direction of the electric field $\langle \Pi_{zz} \rangle$, and orthogonally $\langle\Pi_{\perp z}\rangle = \langle\Pi_{xx}+\Pi_{yy}\rangle{/2}$, all shown in Figure~\ref{elecfieldpressureelectrolyte}.

\begin{figure}
    \centering
    \includegraphics[width=0.9\columnwidth]{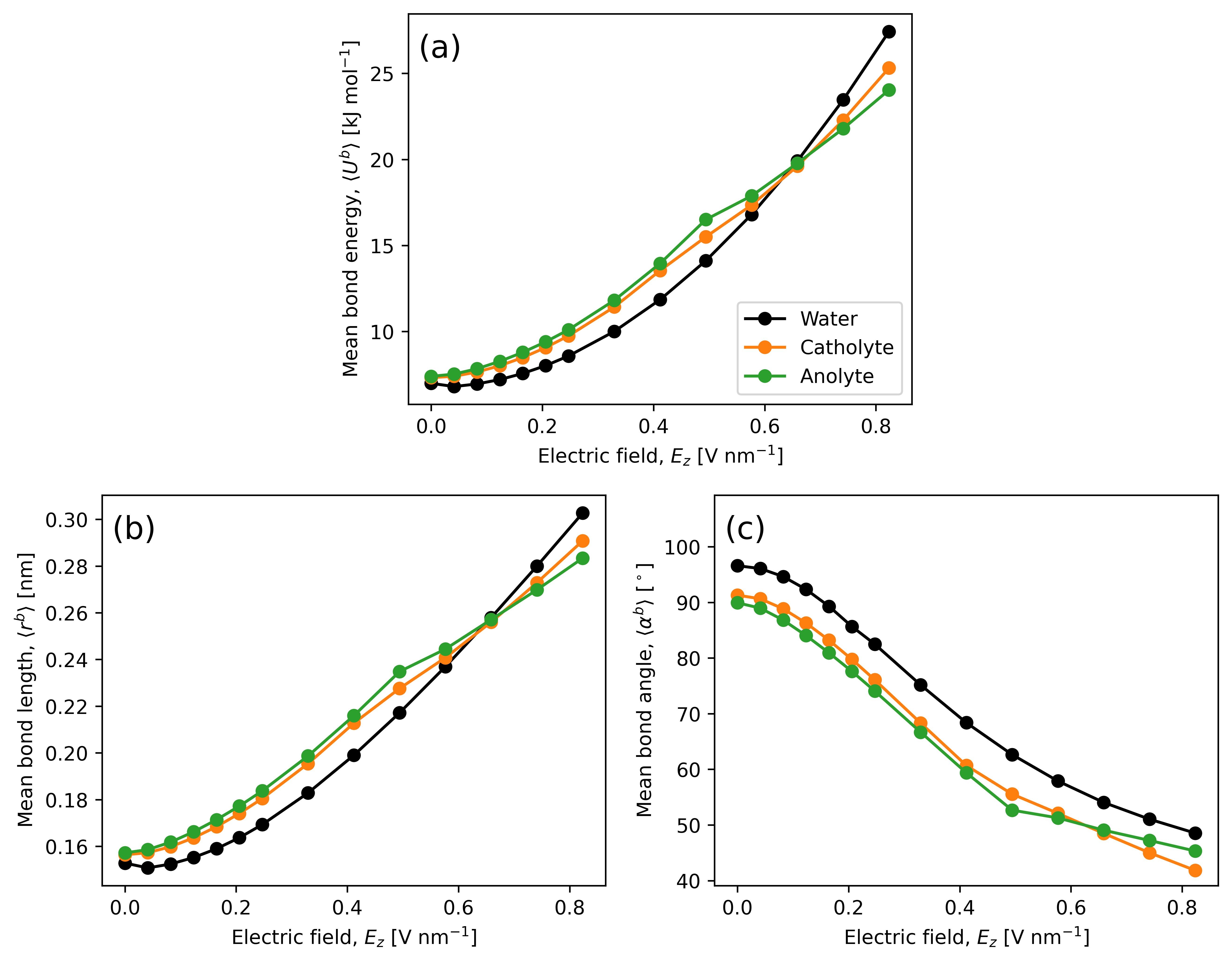}
    \caption{Mean (a) bond energies, (b) bond lengths between central beads and Drude charges, and (c) angles between bonds to Drude charges for each polarisable water entity in pure water (\textbf{black}), catholyte ({\textcolor{orange}{\textbf{orange}}}) and anolyte ({\textcolor{green}{\textbf{green}}}) as functions of induced electric fields.}
    \label{elecfieldbondenergy}
\end{figure}

\begin{figure}
    \centering
    \includegraphics[width=0.8\columnwidth]{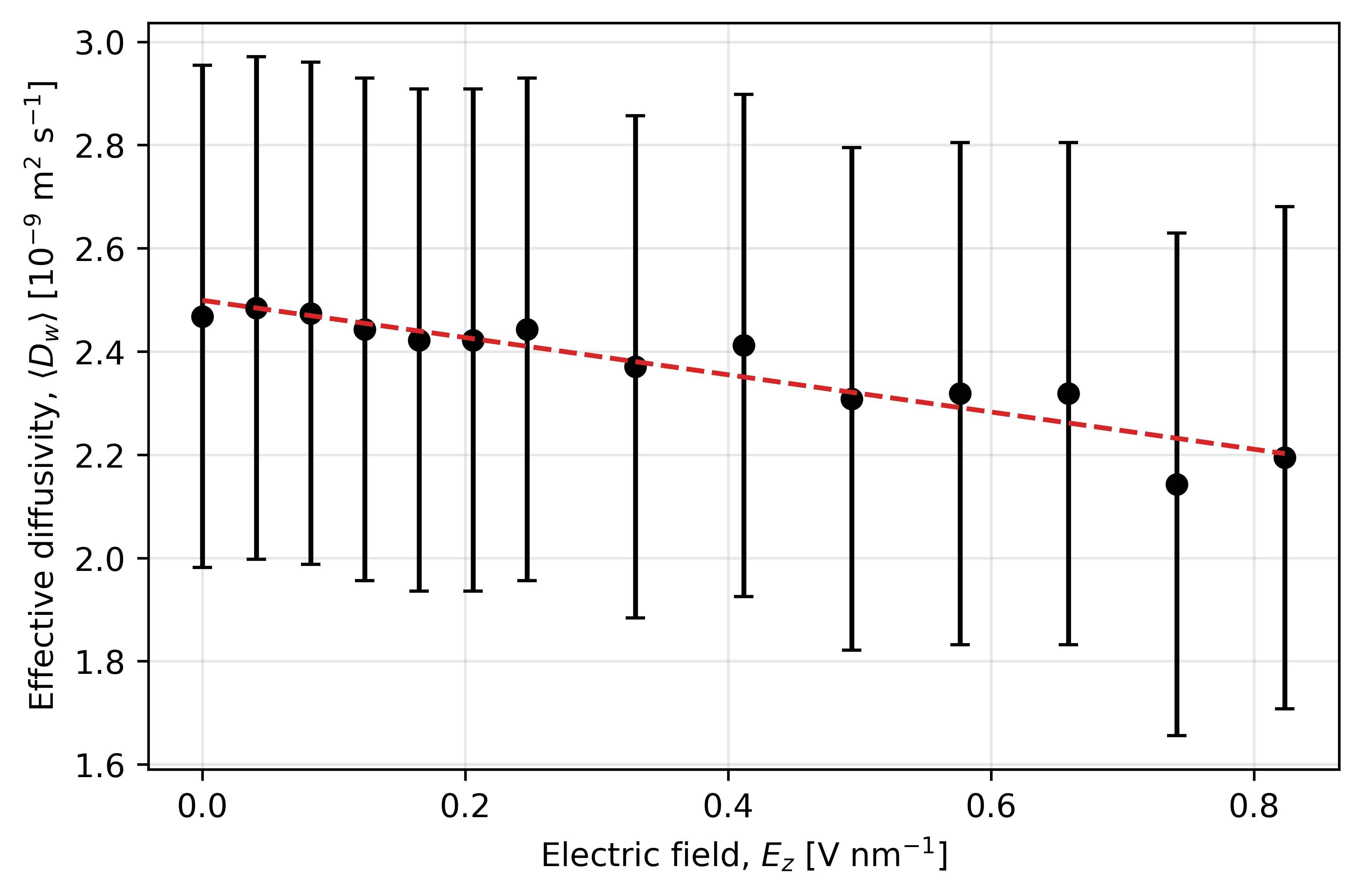}
    \caption{Effective diffusivities of polarisable water, $\langle D_w\rangle$, when subjected to constant electric fields (red dashed line indicating linear best fit: $\langle D_w\rangle [10^{-9}~\textrm{m}^2~\textrm{s}^{-1}] = 2.4990 - 0.3605 E_z [\textrm{V nm}^{-1}]$).} 
    \label{elecfielddiffusionwater}
\end{figure}

Energy components of all main interactions -- short-ranged (\nDPD, Morse), Coulombic (long-ranged) and bonded (topological) -- and thus the configurational energy of the system increase with increasing electric field strength for both the electrolytes and pure water systems.  For these components, the greatest increases are obtained for the bond energy, $\langle U^b\rangle$, between each water central bead and its two Drude satellite charges: these are depicted in Figure~\ref{elecfieldbondenergy}(a) for the two electrolytes and pure polarisable water.  In all three systems, this energy component -- driven by the electric field -- dominates over short-ranged energy increases by around an order of magnitude and those for the Coulombic energy by an additional order.  The kinetic energy deviates from this trend: while it still increases with increasing field strength in the catholyte and anolyte due to generated flows of ions, it barely changes in the pure water system due to the lack of freely-moving ions.  We were able, however, to see a small but discernable downwards trend in the water self-diffusivity with increasing $E_z$, shown in Figure~\ref{elecfielddiffusionwater}. Setting aside the caveat of calculational uncertainties when determining $\langle D_W \rangle$ from MSDs, we note that it is possible for water to become less mobile under very high electric fields, even to the point of freezing into crystalline structures due to enhanced hydrogen bonding \cite{sutmann1998}, and thus a reduction in self-diffusivity at higher $E_z$ should be expected.

\begin{figure}
    \centering
    \includegraphics[width=0.8\columnwidth]{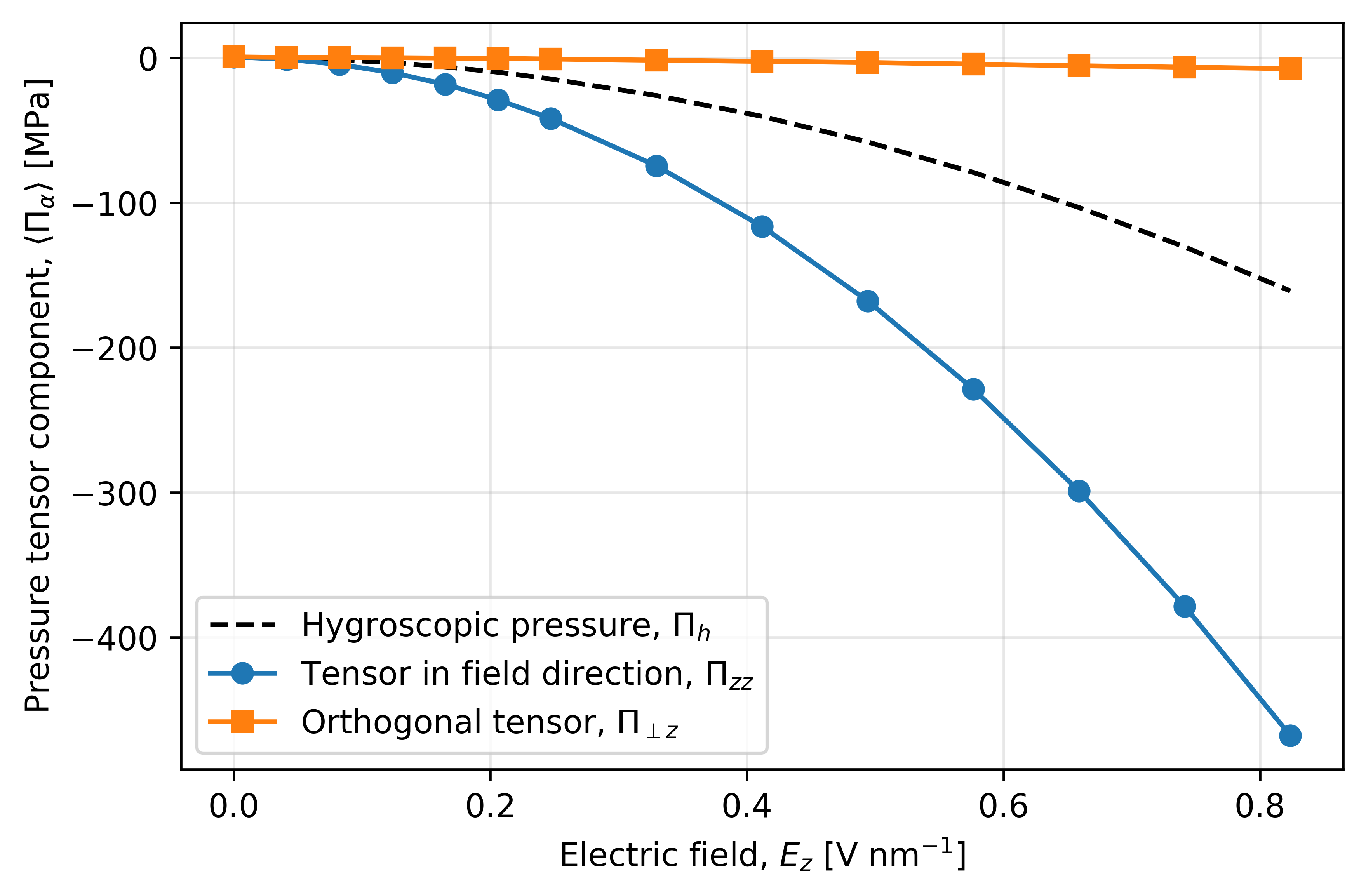}
    \caption{Induced pressure tensor components -- hygroscopic - $\langle\Pi_h\rangle$ dashed \textbf{black} line, in the direction of electric field - $\langle\Pi_z\rangle$ solid {\textcolor{blue}{\textbf{blue}}} line, and orthogonally - $\langle \Pi_{\perp z}\rangle$ solid {\textcolor{orange}{\textbf{orange}}} line -- as functions of constant electric field for polarisable water.}
    \label{elecfieldpressurewater}
\end{figure}

\begin{figure}
    \centering
    \textbf{Catholyte}\\[0.5em]
    \begin{subfigure}{\textwidth}
        \centering
        \includegraphics[width=\textwidth]{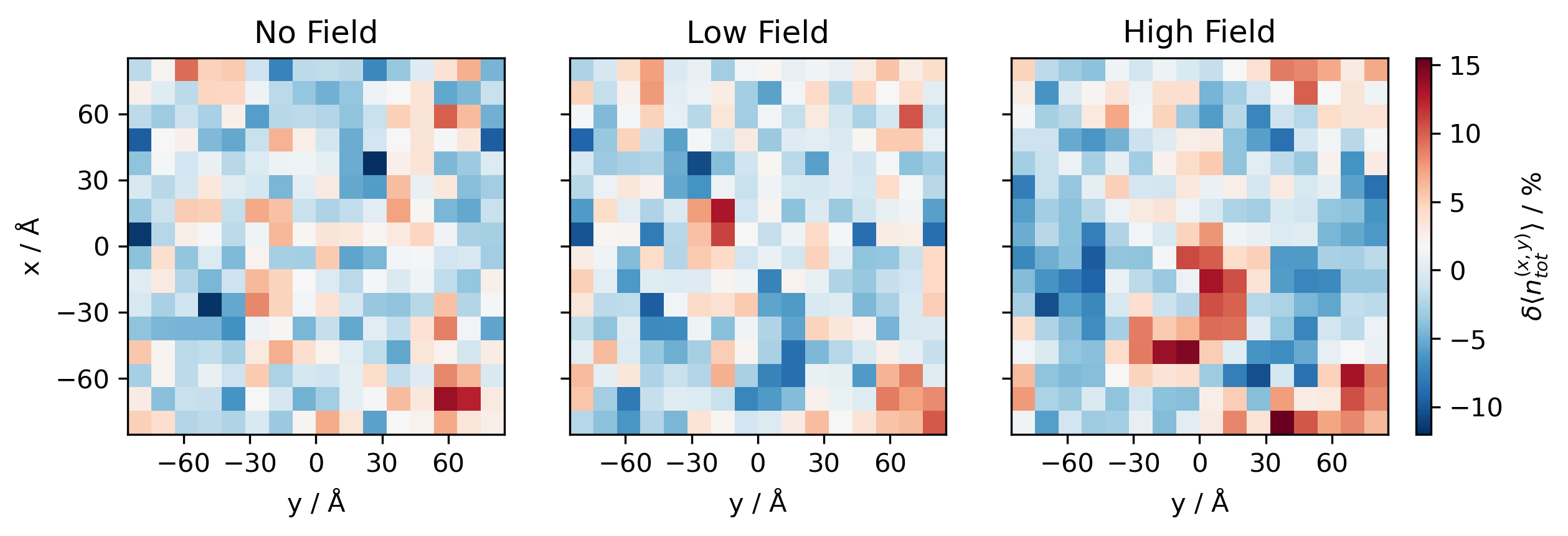}
    \end{subfigure}
    \textbf{Anolyte}\\[0.5em]  
    \begin{subfigure}{\textwidth}
        \centering
        \includegraphics[width=\textwidth]{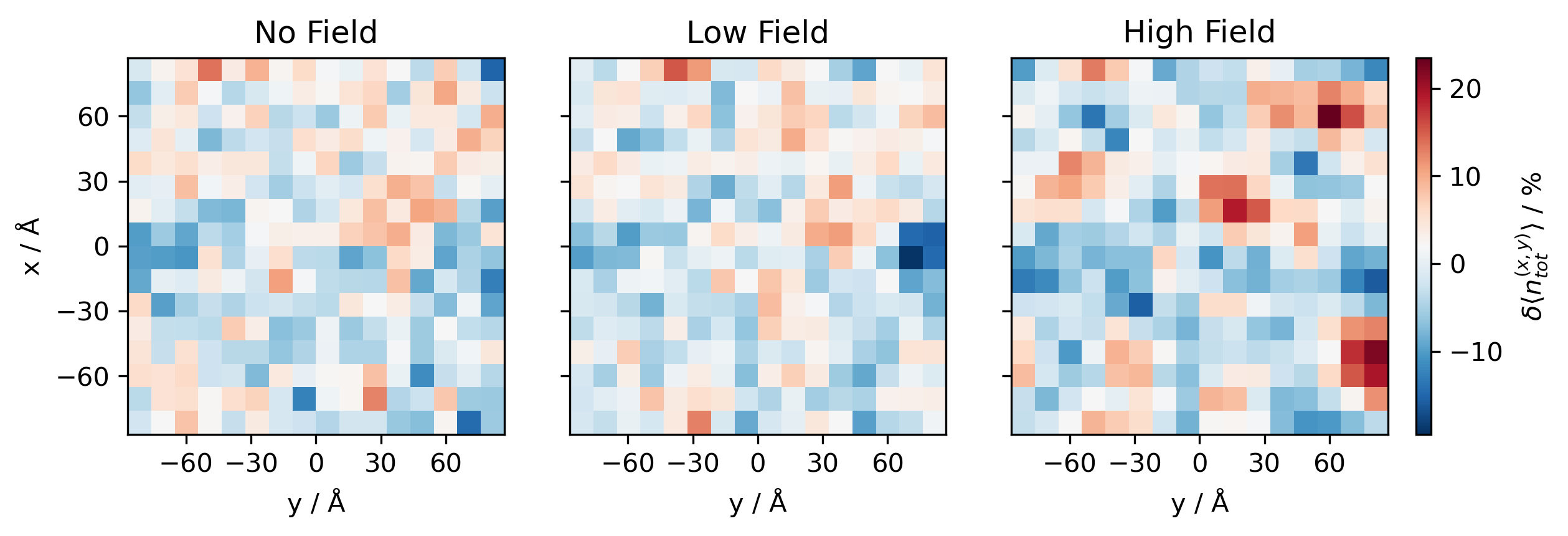}
    \end{subfigure}
    \caption{
        The percentage deviation of the number density of all ionic species (excluding the highly mobile H$^+$ atoms) averaged along the $z$-axis for both the electrolyte systems in the absence of external field and when exposed to low ($E_z=0.0412$~V~nm$^{-1}$) and high ($E_z=0.6588$~V~nm$^{-1}$) external field strengths along the $z$-axis.
    }
    \label{fig:axial_density_grids}
\end{figure}

In Figure~\ref{elecfieldpressurewater} we observe a marked decrease in hygroscopic pressure, $\langle\Pi_h\rangle$, with increasing external field strength in the pure water system. This pressure drop is driven by a well-documented thermodynamic and mechanical phenomenon known as electrostriction \cite{chopra2013} and is proportional to the square of the electric field magnitude:
\begin{equation}
\label{eq:Pelectro}
\Delta \Pi_{\text{el}}=\frac{\epsilon_0}{2}\rho \left(\frac{\partial\epsilon}{\partial\rho}\right)_{T}E^2=\frac{e^2}{8\pi\lambda_B\kT}\gamma_{e}E^2~,
\end{equation}
where $\left(\frac{\partial\epsilon}{\partial\rho}\right)_T$ is the isothermal density derivative of the relative dielectric constant, often coupled into a product with the water density $\rho$ to produce the dimensionless coefficient $\gamma_e = \rho \left(\frac{\partial\epsilon}{\partial\rho}\right)_T$.  Our polarisable water model overestimates $\langle\gamma_e\rangle=53.05$ by an order of magnitude ($\sim 39.5$ times) compared with its experimental value of around $1.35$; this is likely due to the model's \nDPD\ interactions that overestimate the compressibility of water ($\sim 19.7$ times), as seen in Table~\ref{tab:prop}.

It is also worth noting that the pressure orthogonal to the external field $E_z$, $\langle\Pi_{{\perp}z}\rangle$, also decreases with field strength but linearly.  We attribute this effect to greater energetic favourability for radial contraction and axial elongation of the assembly of the water entities -- the two Drude companion beads distancing themselves away from the central bead while also getting closer to each other -- along the dipole axis as the water entities' dipoles increase in magnitude and align along the electric field direction in response to increasing strength.  This is clearly supported in Figure~\ref{elecfieldbondenergy} by changes of the following ensemble averaged properties: (a) embedded bond energy per water entity $\langle U^b\rangle$, (b) Drude bond lengths $\langle r^b\rangle$, and (c) inter-Drude bond angles $\langle\alpha^b\rangle$.  The radial contraction and axial elongation of the water entities into platelets under external field enables the emergence of lamellar-like behaviour, eventually supporting the emergence of liquid rigidification at high external electric field strengths.  

The effect of lamellar-driven liquid rigidifcation is supported by the inhibition of polar water self-diffusivity $D_w$ with increasing external electric field strengths $E_z$ as given in Figure~\ref{elecfielddiffusionwater}.  This inhibition, naturally larger in the orthogonal direction of the external field, essentially enhances the effective self-diffusion of all species in electrolytes along the direction of the external field over that in the perpendicular direction.  In this way it is (i) assisting the flow of ions and of water along the direction of the external field in the electrolytes, as seen in Figure~\ref{elecfielddiffusion}, while also (ii) inhibiting the mixing of electrolyte species and their flows in  perpendicular directions.  These effects ultimately lead to the emergence, splitting and sustenance of external-field aligned, ion-rich and ion-depleted regions in the electrolytes as visualised in Figure~\ref{fig:axial_density_grids}.  This figure depicts percentage variations of number densities for ionic species, $\langle n_i\rangle_z$, with respect to their total average volume density $\overline{n_{tot}}^V_{z}=\sum_{i\in\textrm{ions}}\overline{n}^V_{i,z}$, averaged along the direction of the external field, $z$, on a perpendicularly-laid rectangular grid in $x$ and $y$ directions, at zero, low ($E_z=0.0412$ V nm$^{-1}$) and high field strengths ($E_z=0.6588$ V nm$^{-1}$), sufficiently away from the critical values for each electrolyte, $E_c$.  The plots aim to compare and contrast ensemble-averaged, total ionic density distribution variations (excluding the highly mobile H$^+$ ions):
\begin{equation}
\label{eq:densvar}
\delta\langle n_{tot}^{(x,y)}\rangle=\frac{\langle n_{tot}^{(x,y)}\rangle}{\overline{n}^V_{tot,z}}-1=\frac{\sum_{i\in\textrm{ions}}\langle n_i^{(x,y)}\rangle_z}{{n_{tot}}/{\left({{\mathrm{size}}_x}\cdot{\mathrm{size}}_y\right)}}-1~,
\end{equation}
binned on the $(x,y)$ grids of the same ${\mathrm{size}}_x=16$ and ${\mathrm{size}}_y=16$ over the simulated system volumes for both electrolytes.  The density grid's bounds were chosen systematically to match the maximum and minimum density fluctuations from the high field case for each electrolyte.  The density fluctuation scale for $\delta\langle n_{tot}^{(x,y)}\rangle$ of the anolyte is larger than that of the catholyte because of the availability of ions of higher valency in the anolyte.  However, one can visually discern that the contrast between ion-rich and ion-depleted tiles for equivalent external electric field strengths is always more significant in the catholyte than in the anolyte, consistent with our permitivitty decrement model (Eq.~\ref{eq:pdm}) and confirmed by our results that $\epsilon^{\textrm{cath}}_r<\epsilon^{\textrm{anol}}_r$ prior to the critical values for each electrolyte, $E_c$, (Figure~\ref{elecfieldpolar}).  This is due to a slightly higher overall concentration of ion complexes in the catholyte than in the anolyte (by $\sim 4.5\%$).  At external electric field strengths beyond $E_c$ the availability of ions of higher valency in the anolyte is likely to contribute more to the permitivity decrement and lead to the faster decline of polarisability response, ${\langle P_z\rangle}{/V}$, with $E_z$ than that in the catholyte, reversing the trend at low strengths, as seen in Figure~\ref{elecfieldpolar}.

It can also be shown by decomposing the plots at high field strengths into individual ionic species densities that there is an apparent spatial coupling between certain counter ions, specifically \ce{V^{4+}} (\ce{VO^{2+}}) and \ce{HSO4^{-}} in the catholyte and \ce{V^{3+}} and \ce{HSO4^{-}} in the anolyte.  This represents the grouping of the highest charged ionic species in each electrolyte into coupled counter ion flows.  No other perceptible correlations were observed between the remaining ionic species.

Contrary to the inhibition of water bead self-diffusivity for pure water systems, an 'effective rise' of this property in the electrolytes, $\langle D_w\rangle$, is discernable in Figure~\ref{elecfielddiffusion}.  Although the average water bead dipole moment magnitude, $\langle p_w\rangle$, is discernibly larger in the electrolyte solutions compared with pure water (up until very high $E_z$ strengths), it does not lead to larger dielectric responses or larger permittivities, $\epsilon_r$, for the electrolytes. This is primarily due to a significant drop in polarisable water entity density in the electrolytes: in spite of volume decreases due to the overly high compressibilities of non-polar and polarisable water, several of the latter's entities are replaced with non-polar water beads acting as hydration shells for ions. As such, the overall dipole moment for the system $\langle P \rangle$ decreases, bringing down the relative permittivity as a result.

By comparing Figures~\ref{elecfieldpressureelectrolyte} and \ref{elecfieldpressurewater}, one can see that the behaviour of $\Pi_h$, $\Pi_{z}$, $\Pi_{{\perp}z}$ for our VRFB electrolyte models when subjected to external electric fields, are not dissimilar to those observed in pure solvent at low field strengths.  However, one can clearly see the drastic deviation from the like-water trends, and discern very clearly the critical external field strength points, $E_{cr}^c$ for the catholyte and $E_{cr}^a$ for the anolyte, after which the $\langle P_z\rangle/V$ linear dependence on $E_z$ in Figure~\ref{elecfieldpolar} is broken.

In Figure~\ref{elecfielddiffusion}, it is possible to track changes in behaviour of the effective self-diffusion of elemental species, $\langle D_i\rangle$, in the electrolyte models with increasing external field strengths.  The change in diffusivity regimes, from an approximately quadratic dependence on $E_z$, except for water, to an approximately linear one, occurs at a critical strength, $E_c$, for each electrolyte model.  The increase in the effective self-diffusivity of water $\langle D_w\rangle$ with increasing field strength $E_z$ beyond its value in the absence of external field, $\langle D_w^0\rangle$, is a consequence of Galilean invariance as ensured by the DPD thermostat, i.e. no net centre-of-mass motion is generated during the simulation. Since an applied external electric field generates mass flows of ions along the field's direction or in the opposite direction depending on the ions' charge sign, the solvent must also flow to balance these out. The ion mass flows vary according to their concentrations (mole fractions $x_i$) in mixtures, their charge $q_i$ and mass $m_i$. There are two interesting trends additionally shown in Figure~\ref{elecfielddiffusion}.  The first is a change in effective diffusivities for the two vanadium species in the anolyte with increasing electric field strength: the higher valency ion (\ce{V^{3+}}) becomes more mobile than the lower valency ion (\ce{V^{2+}}), in spite of them being equally mobile when no field is applied. The second is a convergence of effective diffusivities for ion complexes with the same absolute valence but opposite magnitude, particularly (i) \ce{V^{5+}} (\ce{VO2^{+}}) and \ce{HSO4^{-}} in the catholyte, and (ii) \ce{V^{2+}} and \ce{SO4^{2-}} in the anolyte. 

\section{Discussion}

This study generated models of aqueous electrolyte solutions such as those typically found in vanadium redox flow batteries (VRFBs) that operate at the mesoscopic scale (upwards of nanometres and nanoseconds), using a coarse-graining level of 5 \ce{H2O} molecules per modelling entity for water.  Representing both the bulk water in these electrolytes and the accompanying ionic species poses several questions, including -- but not limited to -- how CG entities at these length and time scales should interact, how charge polarisation (particularly of bulk water) should be incorporated, how ionic species in the electrolytes should be represented, and how vastly different dynamic properties for each component can be usefully imposed.

The first two questions above have already been broadly addressed in our previous work on polarising CG water models \cite{seaton2026} but we re-iterate here what is relevant for development of electrolyte models. The choice of short-range (van der Waals-like) interactions between water entities, polarised or not, is key to obtaining correct thermodynamic properties, leading us to choosing \nDPD\ (Eq.~\ref{eq:nDPDpotential}) as an appropriate interaction model for the required scales. Our previous parameterisation of \nDPD\ for water at the desired CG level \cite{seaton2024} was the starting point for both the polarisable water model and for the majority of ions in VRFB electrolytes, given all ions apart from protons are found in complexes with several water molecules each, representing their persistent hydration shell. 

While the \nDPD\ interaction model is an improvement on the typically-used purely repulsive Groot-Warren interaction model \cite{gw97} by offering stable coexistence of two phase systems (e.g. vapour-liquid), the current parameterisations still struggle to correctly obtain all relevant thermodynamic properties at the same time.  Both non-polar and polar water models produce overly high vapour densities and liquid isothermal compressibilities.  While we had at least some prior knowledge of the former property due to expected \nDPD\ phase behaviour \cite{sokhan2023}, the isothermal compressibility was an emergent property -- unlike the Groot-Warren model, where it can be used directly in parameterisation via its known equation of state \cite{gw97} -- and led to higher than expected densities for the VRFB electrolytes, created by combining the polarisable water model with explicit protons and charged non-polar water beads to represent ion-water complexes. In the case of the polar water model, attaching Drude charges to an interacting central bead with harmonic bonds required a reduction in repulsion strength $A_{ij}$ to preserve the liquid density obtained at vapour-liquid equilibrium, although this consequently reduces the interfacial tension.  A potential (albeit extensive) pathway to address these shortcomings would be to re-parameterise \nDPD\ interactions for both water models: for instance, increasing the repulsion power index $n$ and/or reducing the repulsion/attraction parameter $b_{ij}$ might improve fits to interfacial tension and isothermal compressibility.  Our choice to use the \nDPD\ interaction model between polarisable water and ion-water complex beads with an intermediate $A_{ij}$ value is arguably reasonable given the sizes of hydration layers around each ion, although some further improvements could be made to better match thermodynamic properties and to differentiate between anions and cations \cite{nietodraghi2019}.

Beyond thermodynamic properties, the \nDPD\ interaction model additionally provides more realistic hydrodynamic properties, including diffusivities and viscosities, than the Groot-Warren model can ordinarily manage.  On its own (without the DPD thermostat) it can still overestimate diffusivities when used for large coarse-graining levels (see Table \ref{tab:npt}), but the addition of the pairwise DPD thermostat -- Eq. \ref{eq:DPDDissipative} and \ref{eq:DPDRandom} -- provides effective additional control of those properties without disrupting any externally imposed flow fields, which is necessary for simulations at the selected scale and worth a four-fold increase in computational cost. While parameterising the DPD thermostat for the non-polar and polarisable water models to give correct self-diffusivity and dynamic viscosity is comparatively straightforward, doing so for the VRFB electrolytes with disparate diffusivities between ionic species proved to be more challenging due to the need to parameterise between species pairs and different ion and water concentrations in the catholyte and anolyte. Selecting a flexible switching function for the DPD thermostat forces (Eq.~\ref{eq:DPDswitch}) and pragmatic combining rules to determine dissipative force parameters $\gamma_{ij}$ between pairs of unalike species (Eq.~\ref{eq:mixing}) made a parameter search more tractable, producing DPD thermostat parameter sets that can reproduce diffusivities for all species in both VRFB electrolytes that match up well to experimentally-determined values \cite{bromberger2014}. With the exception of protons, it proved advantageous to keep $\gamma_{ij}$ values between pairs of non-alike species comparatively low to avoid over-damping of more mobile species but still non-zero to ensure good temperature control.

To determine the usefulness of the newly-created electrolyte models, we applied external electric fields to them in simulations and studied their behaviours, including comparisons with pure polarisable CG water. Reductions in relative permittivity were observed for the electrolytes compared with water, which can be attributed to their ions acting as polarisability disrupters and affecting the water entities' ability to respond to electric fields. Discontinuities in induced polarisability responses, pressure tensors and effective diffusivities for each species were observed beyond a critical electric field strength for both electrolytes but particularly for the anolyte, indicating a change in ion flow regime that is substantiated by topological changes in water entities, specifically how bond lengths between central beads and Drude satellites and the angles between pairs of bonds vary with increasing field strength.

No such change in behaviour nor any net flow was observed in pure water when electric fields were applied, although a small decrease in water self-diffusivity with increasing electric field strength was detected, indicating the possibility of the polarisable water model being able to represent electrofreezing.  The resulting drop in hygroscopic pressure related to the square of electric field strength also suggests that the same water model can represent electrostriction, and indeed there are observed changes in the shapes of water entities that suggest these deform and elongate along the direction of the field, although its overly high compressibility would need correcting in order for this phenomenon to be correctly represented quantitatively at the chosen coarse-graining level.  We also remark that the water in the electrolytes does not behave in the same manner: increases in its effective diffusivity are observed with increasing electric field strength, in large part due to its need to flow to compensate for mass flow imbalances from ion motion resulting from the same electric field while still obtaining Galilean invariance (i.e. no overall net mass flow).

Analyses of variations of ion densities in the electrolyte systems on a two-dimesional spatial grid perpendicular to an applied external electric field clearly showed the emergence, splitting and sustenance of ion-rich and ion-depleted regions of space over the system volume.  Furthermore, a clear spatial correlation between densities of high-valency counter ions in each electrolyte suggested the establishment of sustained coupled counter-ion flows at high external electric field strengths.  The establishment of high-valence counter-ion coupled flows, together with increasing contrasts between ion-rich and ion-depleted regions in the electrolytes at higher external electric fields beyond the critical strengths, reverses the trend of polarisable response between the catholyte and anolyte from that at lower strengths.

There are a number of shortcomings with the coarse-grained VRFB models we have produced, many of which can be addressed with changes in parameterisation of the underlying \nDPD\ interactions.  However, even in their current form they are adequate for their intended purpose: to generate hydrodynamic properties that are commensurate with experiments and provide for reliable mesoscopic simulations that depend on those properties.  The standard operational conditions of these electrolytes in VRFBs include operating temperatures between $10^{\circ}\textrm{C}$ and $40^{\circ}\textrm{C}$ \cite{ren2023}, internally-generated electric fields of up to $30~\textrm{kV m}^{-1}$ ($3\times10^{-5}~\textrm{V nm}^{-1}$) \cite{wei26,Cecchetti20,yang15,wang26} and flow speeds close to interfaces (including electrodes and felt membranes) less than $0.25~\textrm{m s}^{-1}$ \cite{yadav2023}. Given these conditions, the models' limitations do not prevent them being applied to simulations inside VRFBs, and the fact that they can adequately represent phenomena observed in dielectric materials suggest that the underlying polarisable water model was a suitable choice. 

As such, these highly coarse-grained electrolyte models provide a lot of potential for mesoscopic simulations of electrochemical proceeses. For instance, it is now possible to see how the electrolytes in vanadium redox flow batteries behave at scales and conditions that are normally inaccessible to both experiments and traditional atomistic molecular dynamics simulations.  Insights obtained at these scales could readily be applied as inputs for continuum-scale models of battery operation, providing opportunities to optimise their performance.

\ack{The authors would like to thank our colleagues Silvia Chiacchiera (UKRI STFC Daresbury Laboratory) and Carlos Nieto-Draghi (IFP Energies nouvelles) for useful discussions.}

\funding{This work was supported by the BatCAT consortium, funded under the UK government’s Horizon Europe funding guarantee -- grant number 10112690 -- and by European Union’s Horizon Europe research and innovation programme -- grant agreement number 101137725 -- and made use of computing resources on ARCHER2 provided by CoSeC, the Computational Science Centre for Research Communities, and SCARF by STFC Scientific Computing.}

\roles{Michael A. Seaton - Conceptualization, Methodology, Software, Validation, Formal analysis, Investigation, Writing (Original Draft), Visualization, Supervision.

\noindent{}Benjamin T. Speake - Software, Validation, Formal analysis, Investigation, Writing (Review \& Editing), Visualization.

\noindent{}Tristan Rimmer - Software, Validation, Formal analysis.

\noindent{}Ilian T. Todorov - Conceptualization, Methodology, Validation, Resources, Writing (Review \& Editing), Supervision, Project administration, Funding acquisition.}

\data{The VRFB aqueous electrolyte models are available for download at \href{https://zenodo.org/records/22961058}{https://zenodo.org/records/22961058}.}

\bibliographystyle{iopart-num}
\bibliography{references}

\end{document}